\documentclass[12pt,eadjoint tfnpsf]{article}

\usepackage{amsmath,amsbsy,amssymb,latexsym}

\def\parity{\mathfrak R}

\newcommand{\BE}{\begin{equation}}
\newcommand{\EE}{\end{equation}}
\newcommand{\BEA}{\begin{eqnarray}}
\newcommand{\EEA}{\end{eqnarray}}

\def\12{\frac{1}{2}}
\def\bea{\begin{eqnarray}}
\def\eea{\end{eqnarray}}
\def\ba{\begin{array}}
\def\ea{\end{array}}

\def\one-loop{\mbox{\scriptsize one-loop}}

\def\G{\Gamma}

\catcode`\@=11

\@addtoreset{equation}{section}
\def\theequation{\arabic{section}.\arabic{equation}}

\def\@normalsize{\@setsize\normalsize{15pt}\xiipt\@xiipt
\abovedisplayskip 14pt plus3pt minus3pt%
\belowdisplayskip \abovedisplayskip
\abovedisplayshortskip \z@ plus3pt%
\belowdisplayshortskip 7pt plus3.5pt minus0pt}

\def\small{\@setsize\small{13.6pt}\xipt\@xipt
\abovedisplayskip 13pt plus3pt minus3pt%
\belowdisplayskip \abovedisplayskip
\abovedisplayshortskip \z@ plus3pt%
\belowdisplayshortskip 7pt plus3.5pt minus0pt
\def\@listi{\parsep 4.5pt plus 2pt minus 1pt
\itemsep \parsep
\topsep 9pt plus 3pt minus 3pt}}

\def\underline#1{\relax\ifmmode\@@underline#1\else
$\@@underline{\hbox{#1}}$\relax\fi}
\@twosidetrue

\relax

\catcode`@=12

\catcode`\@=11

\def\section{\@startsection{section}{1}{\z@}{3.5ex plus 1ex minus
.2ex}{2.3ex plus .2ex}{\large\bf}}

\def\thesection{\Roman{section}.}

\def\appendix{\setcounter{section}{0}
\def\thesection{Appendix }

\def\theequation{\Alph{section}.\arabic{equation}}}

\catcode`\@=12

\relax

\def\figcap{\section*{Figure Captions\markboth
{FIGURECAPTIONS}{FIGURECAPTIONS}}\list
{Fig. \arabic{enumi}:\hfill}{\settowidth\labelwidth{Fig. 999:}
\leftmargin\labelwidth
\advance\leftmargin\labelsep\usecounter{enumi}}}
 \relax
\def\tablecap{\section*{Table Captions\markboth
{TABLECAPTIONS}{TABLECAPTIONS}}\list
{Table \arabic{enumi}:\hfill}{\settowidth\labelwidth{Table 999:}
\leftmargin\labelwidth
\advance\leftmargin\labelsep\usecounter{enumi}}}
 \relax
\def\reflist{\section*{References\markboth
{REFLIST}{REFLIST}}\list
{[\arabic{enumi}]\hfill}{\settowidth\labelwidth{[999]}
\leftmargin\labelwidth
\advance\leftmargin\labelsep\usecounter{enumi}}}
 \relax

\newskip\humongous \humongous=0pt plus 1000pt minus 1000pt

\newif\ifdtup

\def\tr{\mathop{\rm tr}}
\def\Tr{\mathop{\rm Tr}}
\def\Im{\mathop{\rm Im}}
\def\Re{\mathop{\rm Re}}

\def\beq{\begin{equation}}
\def\eeq{\end{equation}}

\def\beqn{\begin{eqnarray}}
\def\eeqn{\end{eqnarray}}
\relax

\def\G2{{\; \rm GeV/}c2}
\def\G{\; \rm GeV}

\def\dotx{\dotx{\dot\overline{x}}}

\relax

\newcommand\CL{{\mathcal L}}
\newcommand\CW{{\mathcal W}}
\newcommand\CF{{\mathcal F}}
\newcommand\CD{{\mathcal D}}
\newcommand\fD{{\mathfrak D}}

\newcommand\CN{{\mathcal N}}
\newcommand\I{{\mathbb I}}
\newcommand\CV{{\mathcal V}}
\newcommand\CG{{\mathcal G}}

\def\LLangle{\left< \hspace{-2.5mm} \left<}
\def\RRangle{\right> \hspace{-2.5mm} \right>}
\def\Llangle{\left< \hspace{-1.4mm} \left<}
\def\Rrangle{\right> \hspace{-1.4mm} \right>}
\def\llangle{\left< \hspace{-1mm} \left<}
\def\rrangle{\right> \hspace{-1mm} \right>}

\def\Bar{\overline}

\renewcommand{\thefootnote}{\fnsymbol{footnote}}
\def\fnote#1#2{\begingroup\def\thefootnote{#1}\footnote{#2}\addtocounter
{footnote}{-1}\endgroup}
\begin{document}
%
%
\begin{titlepage}

\begin{flushright}
\normalsize
~~~~
September, 2004 \\
OCU-PHYS 213 \\
hep-th/0409060 \\
\end{flushright}

%
\begin{center}
{\large\bf  Supersymmetric $U(N)$ Gauge Model \\
and Partial Breaking
of ${\cal N}=2$ Supersymmetry }
\end{center}

\vfill

\begin{center}
{%
K. Fujiwara$^a$\footnote{e-mail: fujiwara@sci.osaka-cu.ac.jp}
\quad, \quad
H. Itoyama$^a$\footnote{e-mail: itoyama@sci.osaka-cu.ac.jp}
\quad and \quad
M. Sakaguchi$^b$\footnote{e-mail: msakaguc@sci.osaka-cu.ac.jp}
}
\end{center}

\vfill

\begin{center}
$^a$ \it Department of Mathematics and Physics,
Graduate School of Science\\
Osaka City University\\
\medskip

$^b$ \it Osaka City University Advanced Mathematical Institute
(OCAMI)

\bigskip

3-3-138, Sugimoto, Sumiyoshi-ku, Osaka, 558-8585, Japan \\

\end{center}

\vfill

\begin{abstract}

Guided by the gauging of $U(N)$ isometry associated with the special K\"ahler
 geometry, and the discrete $R$ symmetry, we construct
 the ${\cal N}=2$ supersymmetric action of a $U(N)$ invariant
 nonabelian gauge model in which rigid ${\cal N}=2$
 supersymmetry is spontaneously broken to ${\cal N}=1$.
This generalizes the abelian model considered by Antoniadis,
Partouche and Taylor.
 We shed light on complexity of the supercurrents
  of our model associated with a broken ${\cal N}=2$ supermultiplet of
 currents, and discuss the spontaneously
 broken supersymmetry as an approximate fermionic shift symmetry.

\end{abstract}

\vfill

\setcounter{footnote}{0}
\renewcommand{\thefootnote}{\arabic{footnote}}

\end{titlepage}

\section{Introduction}

Continuing investigations have been made for more than two decades
 on supersymmetric field theories,
 \footnote{See \cite{Wess&Bagger, Weinberg, Sohnius} to review}
 hoping to obtain realistic
 description of nature by broken ${\cal N}=1$ supersymmetry at
 an observable energy scale.
 On the other hand,  it is most natural to view that physics
  beyond this energy scale is controlled by string theory,
   which,  without nontoroidal backgrounds,  produces
 extended supersymmetries in four dimensions.
 Breaking of extended supersymmetries in this vein provides
  a bridge between  gauge field theory and string theory.
String theory does not possess genuine coupling constants:
 instead, they are the vacuum expectation values of 
some supersymmetry preserving moduli fields.
We are thus led to search for the possibility of
 spontaneous partial breaking of extended supersymmetries in
  four dimensions.

 In the context of ${\cal N} =2$ supergravity
\cite{N=2 sugra 1}, spontaneous breaking 
 of local ${\cal N} =2$ supersymmetry to its ${\cal N}=1$ counterpart has
 been accomplished by the simultaneous realization of the Higgs and the
 super Higgs mechanisms.
 Sizable amount of literature has been accumulated till today
 along this direction
\cite{partial susy breaking in N=2 D=4 sugra, AB, David:2003dh}.
 There have been active researches carried out on nonlinear realization
 of extended supersymmetries in the partially broken phase
\cite{HP, AGIT, NLR, NLR 2, NLR 3, Gonzalez-Rey:1998kh}. These are closely
 related to the effective description of string theory \cite{partial susy breaking in string theory}, brane dynamics
\cite{brane actions, brane actions 2, brane actions 3, brane actions 4,
brane actions 5, DeCastro:2004hx} and domain walls \cite{domain wall 4}.
 
 After \cite{HP, AGIT} and prior to the remainder of the works on
 nonlinear realization,  there was a work  within the linear realization done
by  Antoniadis, Partouche and Taylor \cite{APT}
who constructed an
$\mathcal{N}=2$ supersymmetric, self-interacting $U(1)$ model
 with one (or several) abelian $\mathcal{N}=2$ vector multiplet(s) \cite{N=2 constraint} 
  which breaks $\mathcal{N}=2$ supersymmetry
 down to $\mathcal{N}=1$ spontaneously. 
See also \cite{central charge, Ivanov}.
The partial breaking of supersymmetry is accomplished by the
 simultaneous presence of the  electric and  magnetic
  Fayet-Iliopoulos terms, which is a generalization of \cite{FI}.
In the present paper, generalizing the work of \cite{APT}, we construct 
 the ${\cal N}=2$ supersymmetric action of a $U(N)$ invariant
 nonabelian gauge model in which rigid ${\cal N}=2$
 supersymmetry is spontaneously broken to ${\cal N}=1$.
 The gauging of $U(N)$ isometry associated with the special K\"ahler
 geometry, and the discrete $R$ symmetry are  the
 primary ingredients of our construction.

Let us recall that
 partial breaking of extended rigid supersymmetries
 appears not possible on the basis of the positivity of
  the  supersymmetry charge algebra:
\begin{eqnarray}
\left\{ \bar{Q}_{\alpha}^i,Q_{j\dot{\alpha}} \right\}=2(\boldsymbol{1})_{\alpha \dot{\alpha}}\delta_{~j}^{i} H.
\end{eqnarray}
In fact, if $\left. Q_1|0 \right>=0$, one concludes
 $\left. H|0 \right>=0$ and  $\left. Q_i|0 \right> =0$ for all $i$.
If $\left.Q_1|0 \right> \neq 0$,
 then $\left. H|0 \right> =\left. E|0\right>$ with $E>0$
 and $\left. Q_i|0 \right> \neq 0$ for all $i$.
 The loophole to this argument is  that
the use of the local version of the charge algebra is more appropriate
 in spontaneously broken symmetries and the most general supercurrent
 algebra  is 
\begin{eqnarray}
\left\{ \bar{Q}_{\dot{\alpha}}^j,\mathcal{S}_{\alpha i}^m (x) \right\}=2(\sigma^n)_{\alpha \dot{\alpha}} \delta_{~i}^{j} T_n^m(x)+(\sigma^m)_{\alpha \dot{\alpha}} C_i^j, \label{intro}
\end{eqnarray}
where $ \mathcal{S}_{\alpha i}^m $ and  $ T_n^m $ are
 the supercurrents and the energy momentum tensor respectively.
 We have denoted by $C_i^j$
a field independent constant matrix  permitted
 by  the constraints from  the Jacobi identity \cite{Lopuszanski:1978df}.
This last term does not modify the supersymmetry algebra acting on the fields.
 The abelian model of \cite{APT} and our nonabelian generalization
 provide a concrete example of this local algebra
  within linear realization from the point of view of the action principle.

The Lagrangian of our model has  noncanonical kinetic terms coming from
  the nontrivial K\"{a}hler potential and does not fall into the class
 of renormalizable Lagrangians. 
As a model with spontaneously broken 
  $\mathcal{N}=2$ supersymmetry, the prepotential  $\mathcal{F}$ is present
  from the beginning of our construction.
 This is in contrast with breaking $\mathcal{N}=2$ to $\mathcal{N}=1$ 
by the operator (superpotential) $W(\Phi)$, where $\mathcal{F}$ 
appears aposteriori  according to the recent developments
beginning with Dijkgraaf and Vafa \cite{DV}. The model has a $U(1)$
 sector interacting
 with an $SU(N)$ sector  and the spontaneously broken supersymmetry
 acts as an approximate fermionic shift symmetry. Piecing through all
 these properties, we conclude that the action of the model should be
 regarded as a low energy effective action which applies to various
 processes and that the dynamical effects including those of (fractional)
 instantons are to be contained in the prepotential as an input.
 This input should be supplied by a separate means of calculation. 
 The connection with
 the exact determination of the prepotential via \cite{SW, SWC} and
  from integrable systems \cite{SWINT} \cite{DVint}
 offers a new avenue of thoughts with this regard.

In section II, we provide the construction of the
  $\mathcal{N}=2$ supersymmetric action of the $U(N)$ invariant
  nonabelian gauge model which is equipped with
  the Fayet-Iliopoulos $D$ term and a specific superpotential.
 Gauging of the noncanonical kinetic terms coming from
 the K\"{a}hler potential is a necessary step to complete the action.
In section III, we provide the transformation law of the extended
supersymmetries associated with the model. We note that the $SU(2)$
 automorphism of  $\mathcal{N}=2$ supersymmetry   
 has been fixed in the parameter space.
In section IV, we fix the form of the prepotential and  determine
  the vacuum with unbroken gauge symmetry.  We exhibit partial breaking
of $\mathcal{N}=2$ supersymmetry  and discuss a mechanism which
 enables this. 
In section V, we examine a broken $\mathcal{N}=2$
   supermultiplet of currents \cite{IKT} associated with the model.
 The $U(1)_{R}$ current is not conserved except for the
case where the prepotential has an $R$-weight two.
 Despite this, we show that the broken $\mathcal{N}=2$
   supermultiplet of currents provides a useful means to construct
 the extended supercurrents. We shed light upon their complexity.
In section VI, we discuss a role played by the spontaneously broken
 supersymmetry. We see that it acts as a approximate $U(1)$ fermionic
shift symmetry in the limit of letting the magnetic Fayet-Iliopoulos term large 
 relative to the electric one. 
Our discussion in section two and that in section three leading to $\mathcal{N}=2$ supersymmetric Lagrangian exploit an 
algebraic operation denoted by $\parity$.
This operation is defined by including the sign flip of the Fayet-Iliopoulos parameter $\xi \rightarrow -\xi$ into the standard
discrete canonical transformation $R$. It is a legitimate algebraic process to use $\parity$ to demonstrate the second supersymmetry
and in section three we obtain $\mathcal{N}=2$ supersymmetry transformation by demanding the covariance under $\parity$. 
In Appendix A, we give a more pedagogical proof of $\mathcal{N}=2$ supersymmetry of our action, using the canonical $R$.
The two approaches are thus shown to be equivalent. In Appendix B, we reexamine the $\mathcal{N}=1$ current supermultiplet \cite{special Kahler geometry}
in the Wess-Zumino model.


\section{${\CN}=2$ $U(N)$ Gauge Model}

Let us first state our strategy  to obtain  the ${\cal N}= 2$
 supersymmetric action with nonabelian $U(N)$ gauge symmetry.
We adopt the ${\cal N}=1$ superspace formalism to write down  a
 $U(N)$ invariant action consisting of  a set of
 ${\cal N}=1$ $U(N)$ chiral superfields
  and  vector superfields in the adjoint representation.  
The action at this level
 is equipped with  the terms required for the gauging,  
 the Fayet-Iliopoulos
D term, and a generic superpotential. Imposing
the discrete element 
of $SU(2)$ automorphism of ${\cal N}=2$ supersymmety algebra
as symmetry 
of our action  \cite{Weinberg, APT},  we obtain the action
mentioned in the introduction.

What is meant by this last procedure is, however, a little more subtle than
one might first think and we pause to explain this here in more detail.
In the presence of the Fayet-Iliopoulos D term with its coefficient $\xi$,
$\mathcal{N}=1$ Lagrangian is in general not invariant 
under the discrete $R$ symmetry.
 (See (\ref{SU(2):fermion})).  
Best one can do is therefore to consider simultaneously 
an inversion of the parameter $\xi$.
(See (\ref{2.***})). 
Under this extended operation denoted by $\parity$, we will find
\begin{eqnarray}
\parity: \mathcal{L}\to\mathcal{L}, \ \ \ 
\parity: \mathcal{L'}\to \mathcal{L'}.
\end{eqnarray}
(See (\ref{action:off-shell}), (\ref{action:on-shell}).) 
Combining this with the algebra
\begin{eqnarray}
\parity \delta_1 \parity^{-1} =\delta_2, \label{parityalgebra}
\end{eqnarray}
we conclude that our final actions (\ref{action:on-shell}) 
and (\ref{off-shellaction2}) with (\ref{keyrelation1}) and 
(\ref{psi-lambda}) are invariant 
under $\mathcal{N}=2$ supersymmetry. 
Here we denote by $\delta_1$ and $\delta_2$, 
the transformation of the first supersymmetry 
and that of the second supersymmetry respectively. 
This definition $\parity$ turns out to be consistent 
with an interpretation that full rigid $SU(2)$ symmetry has been 
fixed in the parameter space. 
This is discussed in section III.

\subsection{$U(N)$ Gauge Model}

Let us introduce a set of $\CN=1$ chiral superfields
\begin{eqnarray}
\Phi(x^m,\theta) &=& \sum_{a=0}^{N^2-1} \Phi ^a t_a~.
\end{eqnarray}
Here,
$t_a$, $a=0, 1,\ldots ,(N^2-1)$, are $N\times N$ hermitian matrices
which generate $u(N)$ algebra,
and $t_{\hat a}$, $\hat a=1,\ldots ,(N^2 -1)$, generate $su(N)$ algebra
\begin{eqnarray}
[t_{\hat a}, t_{\hat b}]=if^{\hat c}_{\hat a\hat b}t_{\hat c}.
\end{eqnarray}
The index $0$ refers to the overall $u(1)$ generator.
The scalar fields $A=A^at_a$ in $\Phi$
undergo the adjoint action
\begin{eqnarray}
A\to UAU^\dagger,
\end{eqnarray}
under $U(N)$.

The kinetic term for $A$ is generated by
\begin{eqnarray}
\CL_{K}&=&
\int d^2\theta d^2\bar\theta ~K(\Phi^a,\Phi^{*a}),
\end{eqnarray}
where $K(A^a, A^{*a})$ is the K\"ahler potential.
The K\"ahler potential we employ is given by
\begin{eqnarray}
& & K(A^a, A^{*a})=\frac{i}{2} (A^a \CF^*_a
- A^{*a} \CF_a),
\label{Kahler potential}
\end{eqnarray}
where $\CF_a=\partial_a\CF=\frac{d}{dA^a}\CF$
and $\CF$ is an analytic function of $A$.\fnote{$\star$}{
The $\Omega=({A^a \atop \CF_b})$ can be regarded as
a section of a holomorphic symplectic bundle
on a special K\"ahler geometry
(see \cite{special Kahler geometry}
and references therein).
We work in special coordinates in this paper.
}
The K\"ahler potential can be written using a
hermitian metric on the bundle
compatible with the symplectic structure
 as
\begin{eqnarray}
K=-\frac{i}{2}\left\langle\Omega | \bar\Omega\right\rangle ,~~~~
\left\langle\Omega | \bar\Omega\right\rangle=-\Omega^T
\left(
\begin{array}{cc}
0 & \I \\
-\I & 0 \\
\end{array}
\right)\Omega^* ~.
\end{eqnarray}
The K\"ahler metric
\begin{eqnarray}
g_{ab^*}=\partial_a\partial_{b^*}K=
\Im \CF_{ab}
\end{eqnarray}
constructed this way always admits
a $U(N)$ isometry.
The holomorphic Killing vectors
$k_a=k_a{}^b\partial_b$ are
generated by the Killing potential $\fD_a$,
to be introduced shortly,
as
\begin{eqnarray}
k_a{}^b=-ig^{bc^*}\partial_{c^*}\fD_a,~~~
k_a^*{}^{b}=ig^{cb^*}\partial_{c}\fD_a.
\end{eqnarray}
These form an algebra $
[k_a,k_b]=-f^c_{ab}k_c.
$
The $A^a$ and $\CF_a$ transform in the adjoint representation
of $U(N)$
\begin{eqnarray}
&&
\delta_bA^a=-f^a_{bc}A^c,~~~~
\delta_b\CF_a=-f^c_{ab}\CF_c ~.
\label{adjoint tfn}
\end{eqnarray}
One finds that the commutator of two of $\delta_a$ is
given by
$
[\delta_a,\delta_b]=f_{ab}^c\delta_c.
$
Comparing this with the commutator of two Killing vectors,
we are able to identify $\delta_a$ with $-k_a$.
The equation (\ref{adjoint tfn}) is rewritten as
\begin{eqnarray}
k_b{}^c\partial_cA^a=f^a_{bc}A^c,~~~~
k_b{}^c\partial_c\CF_a=-f^c_{ba}\CF_c ~.
\label{adjoint}
\end{eqnarray}
The isometry group can be embedded in the symplectic group,
and the $\fD_a$ is given by
\begin{eqnarray}
\fD_a=
-\frac{1}{2}\left\langle\Omega | T_a \bar\Omega\right\rangle
=
-\frac{1}{2}(\CF_bf^b_{ac} A^{*c}+\CF_{b}^*f^b_{ac}A^c) ,~~~~
T_a=\left(
\begin{array}{cc}
f^b_{ac} & 0 \\
0 & -f^b_{ac} \\
\end{array}
\right).
\label{Killing potential}
\end{eqnarray}
Note that $\fD_{\hat{a}}$ are completely determined by this formula
while  $\fD_0$ is determined up to a constant.

In order to gauge the $U(N)$ isometry, we introduce
a set of $\CN=1$ vector superfields
\begin{eqnarray}
V(x^m,\theta,\bar{\theta})&=& \sum_{a=0}^{N^2-1} V^a t_a.
\end{eqnarray}
The $U(N)$ gauging of $\CL_K$ is accomplished \cite{gauging} by adding 
\begin{eqnarray}
\CL_\Gamma=\int d^2\theta d^2\bar\theta \Gamma,~~~~
\Gamma=
\left[
\int^1_0d\alpha e^{\frac{i}{2}\alpha v^a(k_a-k_a^*)}v^c\fD_c
\right]_{v^a\to V^a},
\end{eqnarray}
where $[\cdots ]_{v^a\to V^a}$ means the replacement of $v^a$ by $V^a$
after evaluating $\cdots $.
Combining $\CL_{K}$ with $\CL_\Gamma$,
we obtain
\begin{eqnarray}
\CL_K+\CL_\Gamma&=&
-g_{ab^*}\CD_mA^a\CD^mA^{*b}
-\frac{i}{2}g_{ab^*}\psi^a\sigma^m\CD_m'\bar\psi^b
+\frac{i}{2}g_{ab^*}\CD_m'\psi^a\sigma^m\bar\psi^b
\nonumber\\&&
+g_{ab^*}F^aF^{*b}
-\frac{1}{2}g_{ab^*,c^*}F^a\bar\psi^b\bar\psi^c
-\frac{1}{2}g_{bc^*,a}F^{*c}\psi^a\psi^b
\nonumber\\&&
+\frac{1}{\sqrt{2}}g_{ab^*}(\lambda^c\psi^ak_c^*{}^{b}
+\bar\lambda^c\bar\psi^bk_c{}^{a})
+\frac{1}{2}D^a\fD_a~,
\label{L;K}
\end{eqnarray}
where we have
exploited 
$\frac{1}{4}g_{ac^*,bd^*}\psi^a\psi^b\bar\psi^c\bar\psi^d=0$
as $g_{ac^*,bd^*}=0$
for the choice of $K$ in (\ref{Kahler potential}).
The covariant derivatives are defined as
\begin{eqnarray}
\CD_mA^a&=&
\partial_mA^a
-\frac{1}{2}v_m^bk_b{}^a,\\
\CD_m'\psi^a&=&
\CD_m\psi^a+\Gamma^a_{bc}\CD_mA^b\psi^c,\\
\CD_m\psi^a&=&
\partial_m\psi^a
-\frac{1}{2}v_m^b\partial_ck_b{}^a\psi^c~,
\label{D psi}
\end{eqnarray}
where $\Gamma^a_{bc}=g^{ad^*}g_{bd^*,c}$.

The gauged kinetic action for the vector superfield $V$
is given by
\begin{eqnarray}
\CL_{\CW^2}&=&
-\frac{i}{4}\int d^2\theta \tau_{ab}\CW^a\CW^b + c.c. ~,~~~~
{\mathcal W}_{\alpha}=-\frac{1}{4} \bar{D} \bar{D} e^{-V} D_{\alpha}
e^V={\mathcal W}_{\alpha}^a t_a~,
\end{eqnarray}
where
$
\tau_{ab}
=(\tau_1)_{ab}+i (\tau_2)_{ab}
$ is an analytic function of $\Phi$,
and will be determined by requiring $\CN=2$ supersymmetry.
The $\CL_{\CW^2}$ is evaluated as
\begin{eqnarray}
\CL_{\CW^2}&=&
-\frac{1}{2}\tau_{ab}\lambda^a\sigma^m\CD_m\bar\lambda^b
-\frac{1}{2}\bar\tau_{ab}\CD_m\lambda^a\sigma^m\bar\lambda^b
-\frac{1}{4}(\tau_2)_{ab}v_{mn}^av^{bmn}
-\frac{1}{8}(\tau_1)_{ab}\epsilon^{mnpq}v_{mn}^av_{pq}^b
\nonumber\\&&
-i\frac{\sqrt{2}}{8}(
\partial_c\tau_{ab}
\psi^c\sigma^n\bar\sigma^m\lambda^a
-\partial_{c^*}\tau^*_{ab}
\bar\lambda^a\bar\sigma^m\sigma^n\bar\psi^c
)v_{mn}^b
\nonumber\\&&
+\frac{1}{2}(\tau_2)_{ab}D^aD^b
+\frac{\sqrt{2}}{4}(
\partial_c\tau_{ab}\psi^c\lambda^a
+\partial_{c^*}\tau^*_{ab}\bar\psi^c\bar\lambda^a
)D^b
+\frac{i}{4}\partial_c\tau_{ab}F^c\lambda^a\lambda^b
-\frac{i}{4}\partial_{c^*}\tau^*_{ab}F^{*c}\bar\lambda^a\bar\lambda^b
\nonumber\\&&
-\frac{i}{8}\partial_c\partial_d\tau_{ab}\psi^c\psi^d\lambda^a\lambda^b
+\frac{i}{8}\partial_{c^*}\partial_{d^*}\tau^*_{ab}
\bar\psi^c\bar\psi^d\bar\lambda^a\bar\lambda^b,
\label{L:gauge}
\end{eqnarray}
where we have defined
\begin{eqnarray}
v_{mn}^a&=&
\partial_mv_n^a
-\partial_nv_m^a
-\frac{1}{2}f^a_{bc}v^b_mv^c_n,\\
\CD_m\lambda^a&=&
\partial_m\lambda^a
-\frac{1}{2}f^a_{bc}v_m^b\lambda^c.
\label{D lambda}
\end{eqnarray}

In addition, we include the superpotential term
\begin{eqnarray}
\CL_W&=&
\int d^2\theta W(\Phi) +c.c.
\nonumber \\&=&
F^a\partial_a W
-\frac{1}{2}\partial_a\partial_b W\psi^a\psi^b
+F^{*a}\partial_{a^*}  W^*
-\frac{1}{2}\partial_{a^*}\partial_{b^*}  W^*\bar\psi^a\bar\psi^b ~,
\label{L:superpotential}
\end{eqnarray}
and the Fayet-Iliopoulos D-term \cite{FI}
\begin{eqnarray}
\CL_{D}&=&  \xi \int d^2 \theta d^2 \bar{\theta} V^0 =
 \sqrt{2}\xi D^0.
\end{eqnarray}
The superpotential $W$ will be determined by requiring $\CN=2$ supersymmetry.
Finally,
putting all these together,
the total action is given as
\begin{eqnarray}
\CL=\CL_K+\CL_\Gamma+\CL_{\CW^2}+\CL_W+\CL_D.
\label{action:off-shell}
\end{eqnarray}

For the sake of our discussion in the next subsection,
we present the on-shell action, eliminating the auxiliary fields by using the equations
of motion

\begin{eqnarray}
D^{a}&=&
\hat{D}^a
-(\tau_2^{-1})^{ab}\left(
\frac{1}{2}\fD_b + \sqrt{2}\xi\delta_{b}^0
\right),\label{2.23}
\\
F^a&=&\hat{F}^a
-g^{ab^*}\partial_{b^*} W^*,
\\
F^{*a}&=&\hat{F}^{*a}
-g^{ba^*}\partial_{b} W~,\label{2.25}
\end{eqnarray}
where
\begin{eqnarray}
\hat{D}^a&=&-\frac{\sqrt{2}}{4}(\tau_2^{-1})^{ab} \left( \partial_d\tau_{bc}\psi^d\lambda^c+\partial_{d^*}\tau^*_{bc}\bar\psi^d\bar\lambda^c \right), \label{Dhat} \\
\hat{F}^a&=&-g^{ab^*} \left( -\frac{i}{4}\partial_{b^*}\tau^*_{cd}\bar\lambda^c\bar\lambda^d-\frac{1}{2}g_{cb^*,d}\psi^c\psi^d \right), \label{Fhat} \\
\hat{F}^{*a}&=&-g^{ba^*} \left( \frac{i}{4}\partial_{b}\tau_{cd}\lambda^c\lambda^d-\frac{1}{2}g_{bc^*,d^*}\bar\psi^c\bar\psi^d \right). \label{Fhat*}
\end{eqnarray}
The action $\CL$ takes the following form;
\begin{eqnarray}
\label{action:on-shell}
\CL'
&=&
\CL_{\rm{kin}}
+\CL_{\rm{pot}}
+\CL_{\rm{Pauli}}
+\CL_{\rm{mass}}
+\CL_{\rm{fermi^4}}
\end{eqnarray}
where
\begin{eqnarray}
\CL_{\rm{kin}}&=&
-g_{ab^*}\mathcal{D}_m A^a \mathcal{D}^m A^{*b}
-\frac{1}{4} (\tau_2)_{ab} v_{mn}^a v^{bmn}
-\frac{1}{8} (\tau_1)_{ab} \epsilon^{mnpq}v_{mn}^a v_{pq}^b
\\&&
-\frac{1}{2} \tau_{ab} \lambda^a \sigma^m \mathcal{D}_m
\bar{\lambda}^b
-\frac{1}{2} {\tau}^*_{ab} \mathcal{D}_m \lambda^a
\sigma^m \bar{\lambda}^b
-\frac{i}{2}g_{ab^*} \psi^a\sigma^m \mathcal{D}_m \bar{\psi}^b
+\frac{i}{2}g_{ab^*} \mathcal{D}_m \psi^a \sigma^m \bar{\psi}^b,
\nonumber \\
\CL_{\rm{pot}}&=&
-\frac{1}{2} \left( \tau_2^{-1} \right)^{ab} \left(
\frac{1}{2}\fD_{a} + \sqrt{2} \xi\delta_{a}^0
\right)
\left(
\frac{1}{2}\fD_{b} + \sqrt{2} \xi\delta_{b}^0
\right)
-g^{ab^*}\partial_a W \partial_{b^*} {W}^*,
\\
\CL_{\rm{Pauli}}&=&
-i\frac{\sqrt{2}}{8} \partial_c \tau_{ab} \psi^c \sigma^n
\bar{\sigma}^m \lambda^a v_{mn}^b
+i\frac{\sqrt{2}}{8} \partial_{c^*}
{\tau}^*_{ab} \bar{\lambda}^a \bar{\sigma}^m \sigma^n
\bar{\psi}^cv_{mn}^b,
\\
\CL_{\rm{mass}}&=&
-\frac{1}{2}\partial_a\partial_b W \psi^a \psi^b
-g^{ab^*} \partial_a W \left(
-\frac{i}{4}\partial_{b^*} {\tau}^*_{cd} \bar{\lambda}^c
\bar{\lambda}^d
-\frac{1}{2} g_{cb^*,d} \psi^c \psi^d
\right)
\nonumber \\&&
-\frac{1}{2}\partial_{a^*} \partial_{b^*} {W}^*
\bar{\psi}^a \bar{\psi}^b
-g^{ab^*} \left(
\frac{i}{4} \partial_a \tau_{cd} \lambda^c \lambda^d
-\frac{1}{2}g_{ac^*,d^*} \bar{\psi}^c \bar{\psi}^d
\right) \partial_{b^*} {W}^*
\nonumber \\&&
+\frac{1}{\sqrt{2}} g_{ab^*} \left( \bar{\lambda^c} \bar{\psi}^b
k_c{}^a
+\lambda^c \psi^a k_c^*{}^{b} \right)
\nonumber \\&&
-\frac{\sqrt{2}}{4} \left( \tau_2^{-1} \right)^{ab}
\left(\frac{1}{2}\fD_{a} + \sqrt{2} \xi\delta_{a}^0 \right)
\left(
\partial_d \tau_{bc}\psi^d \lambda^c
+\partial_{d^*}{\tau}^*_{bc} \bar{\psi}^d \bar{\lambda}^c
\right),
\\
\CL_{\rm{fermi^4}}&=&
-\frac{i}{8}\partial_c \partial_d \tau_{ab}
\psi^c \psi^d \lambda^a\lambda^b
+\frac{i}{8} \partial_{c^*} \partial_{d^*}
{\tau}^*_{ab}\bar{\psi}^c\bar{\psi}^d\bar{\lambda}^a\bar{\lambda}^b
\nonumber \\&&
-\frac{1}{16}\left(\tau_2^{-1} \right)^{ab}
\left( \partial_d \tau_{ac} \psi^d \lambda^c
+\partial_{d^*} {\tau}^*_{ac} \bar{\psi}^d \bar{\lambda}^c
\right)
\left( \partial_f \tau_{be} \psi^f \lambda^e
+\partial_{f^*} {\tau}^*_{be} \bar{\psi}^f \bar{\lambda}^e
\right)
\nonumber \\&&
-g^{ab^*}
\left(
\frac{i}{4}\partial_a \tau_{cd} \lambda^c\lambda^d
-\frac{1}{2}g_{ac^*,d^*}\bar{\psi}^c \bar{\psi}^d
\right)
\left(
-\frac{i}{4}\partial_{b^*}{\tau}^*_{ef}
\bar{\lambda}^e\bar{\lambda}^f
-\frac{1}{2}g_{eb^*,f} \psi^e \psi^f
\right).
\end{eqnarray}

\subsection{Discrete  $R$-symmetry}

We shall show that our  Lagrangian (\ref{action:on-shell}) 
$\mathcal{L'}$
can be made
invariant
under the action
\begin{eqnarray}
\parity: \left(
  \begin{array}{c}
    \lambda^a   \\
    \psi^a   \\
  \end{array}
\right)
\to
\left(
  \begin{array}{c}
    \psi^a  \\
    -\lambda^a   \\
  \end{array}
\right)~,
\label{SU(2):fermion}
\end{eqnarray}
which is a discrete element of the $SU(2)$ $R$-symmetry 
that acts as  an automorphism
of $\CN=2$ supersymmetry.

First, we examine the invariance of
$\CL_{\rm{Pauli}}$, $\CL_{\rm{fermi^4}}$ and $\CL_{\rm{kin}}$
 under the action (\ref{SU(2):fermion}).
The invariance of $\CL_{\rm{Pauli}}$ and
that of 
$\CL_{\rm{fermi^4}}$
under (\ref{SU(2):fermion})
 require
\begin{eqnarray}
\partial_c\tau_{ab}=\partial_a\tau_{cb},~~~
\end{eqnarray}
and
\begin{eqnarray}
\partial_c\partial_d\tau_{ab}=\partial_a\partial_b\tau_{cd}~,~~
\partial_c\tau_{ab}=\CF_{abc},
\label{condition:Pauli}
\end{eqnarray}
respectively.
In addition,
the invariance of the
fermion kinetic terms in
$\CL_{\rm{kin}}$
implies that
\begin{eqnarray}
\Im(\tau_{ab})=\Im(\CF_{ab})
\label{condition:fermion kinetic:tau}
\end{eqnarray}
and
\begin{eqnarray}
-2\partial_a\partial_{b^*}\fD_c=
\tau_{ad}f^d_{cb}+\tau^*_{bd}f^d_{ca},
\label{condition:fermion kinetic:KP}
\end{eqnarray}
as well as the last condition in (\ref{condition:Pauli})
which comes from that the terms with a derivative of $A^*$ vanish.
The first condition (\ref{condition:fermion kinetic:tau})
comes from
the terms with a derivative of $\lambda$ or $\psi$
while the second one (\ref{condition:fermion kinetic:KP})
from those including $v_m^a$. 
For
the boson kinetic terms in $\CL_{\rm{kin}}$, 
the invariance is obvious
because they do not contain fermionic fields. From
the conditions
(\ref{condition:Pauli}) and
(\ref{condition:fermion kinetic:tau}),
we conclude that
\begin{eqnarray}
\tau_{ab}=\CF_{ab},
\label{tau}
\end{eqnarray}
so that $g_{ab^*}=(\tau_2)_{ab}$.
It is easy to show that
the Killing potential
$\fD_a$ defined in (\ref{Killing potential}) solves
the condition (\ref{condition:fermion kinetic:KP}).

Secondly,
we examine the invariance of the $\lambda\lambda$ and $\psi\psi$
mass terms in $\CL_{\rm{mass}}$
under (\ref{SU(2):fermion}).
The key relation
required 
 for this invariance is
\begin{eqnarray}
-\frac{i}{4}g^{cd^*}\partial_c\tau_{ab}\partial_{d^*} W^* =
\frac{1}{2}g^{cd^*}\partial_cWg_{ad^*,b}
-\frac{1}{2}\partial_a\partial_b W. \label{keyrelation1}
\end{eqnarray}
Writing the $U(N)$ invariant function $W$
as
$W=eA^0+m\phi(A)$, where the $e$ and $m$ are real constants,
it reduces to
\begin{eqnarray}
\CF_{abc}(\frac{1}{\CF-\CF^*})^{cd}
(\partial_d\phi-\partial_{d^*}\phi^*)
=\partial_a\partial_b\phi,
\end{eqnarray}
which can be solved by $\phi=\CF_0+ const.$
Thus we can choose
\begin{eqnarray}
W=eA^0+m\CF_0,
\label{W}
\end{eqnarray}
up to an irrelevant constant.

Thirdly,
we examine the $\psi\lambda$ terms in $\CL_{\rm{mass}}$.
Because $\psi^a\lambda^b$ is odd under the action
(\ref{SU(2):fermion}), the coefficient,
$\frac{1}{\sqrt{2}}g_{ac^*}k_b^*{}^{c}-\frac{\sqrt{2}}{8}(\tau_2^{-1})^{cd}\partial_a\tau_{cb}
(\fD_d+2\sqrt{2}\xi\delta_{d}^0)
$, must be odd.
This implies the key relation for the invariance
\begin{eqnarray}
i\partial_a\fD_b+i\partial_b\fD_a
-\frac{1}{2}(\tau_2^{-1})^{cd}\partial_a\tau_{cb}
\fD_{d}
=0,
\label{psi-lambda}
\end{eqnarray}
as well as
\begin{eqnarray}
\parity ; \ \xi \rightarrow - \xi. \label{2.***}
\end{eqnarray}
The equation (\ref{psi-lambda}) can be proven as follows.
First, we note that
\begin{eqnarray}
\CF_{ac}f^c_{db}+\CF_{bc}f^c_{da}=-\CF_{abc}f^c_{de}A^e,
\end{eqnarray}
which is derived as a derivative of
the second relation in (\ref{adjoint}).
Using this relation and the definition (\ref{Killing potential}), one finds that
\begin{eqnarray}
i\partial_a\fD_b+i\partial_b\fD_a=
-\frac{i}{2}\CF_{abc}f^c_{de} A^{*d}A^e.
\label{deform:dD}
\end{eqnarray}
On the other hand, the Killing potential is shown to be
rewritten as
\begin{eqnarray}
\fD_a=\frac{1}{2}f^b_{cd} A^{*c}A^d(\CF^*_{ab}-\CF_{ab})
=-ig_{ab}f^b_{cd} A^{*c}A^d
\label{deform:KP}
\end{eqnarray}
by using the second relation in (\ref{adjoint}).
The equations (\ref{deform:dD}) and (\ref{deform:KP})
are enough to see that the equation
(\ref{psi-lambda}) is true.

Lastly, we examine $\CL_{\textrm{pot}}$.
The invariance of $\CL_{\textrm{pot}}$ under (\ref{2.***})
follows from the
fact that the term linear in $\xi$ in $\CL_{\textrm{pot}}$
vanishes:
\begin{eqnarray}
-\frac{1}{2}(\tau_2^{-1})^{ab}\fD_a\sqrt{2}\xi\delta_{b}^0
=-\frac{\sqrt{2}}{2}\xi g^{a0}(-ig_{ab}f^b_{cd} A^{*c}A^d)
=i\frac{\sqrt{2}}{2}\xi
f^0_{cd} A^{*c}A^d=0
\label{xi linear}
\end{eqnarray}
where we have used (\ref{tau}) and (\ref{deform:KP}).

In summary, we have shown that our on-shell action
(\ref{action:on-shell})
admits the discrete
$\parity$-symmetry (\ref{SU(2):fermion})
and (\ref{2.***})
 if we choose $\tau_{ab}$ as (\ref{tau})
and $W$ as (\ref{W}).

We will show that the discrete $\parity$-symmetry
can be realized in the off-shell action (\ref{action:off-shell})
with (\ref{tau}) and (\ref{W}).
In an ungauged theory without a superpotential,
the discrete action on the auxiliary fields is
$D^a\to -D^a$ and $F^a\to F^{*a}$.
In our model, this is modified as is seen below.
The terms which need to be checked are those including
auxiliary fields.
First, we examine bosonic terms including $F^a$ and $F^{*a}$,
\begin{eqnarray}
g_{ab^*}F^aF^{*b}+F^a\partial_aW+F^{*a}\partial_{a^*} W^*.
\end{eqnarray}
Apparently, this is not invariant under $F\to F^*$.
Rewriting it as
\begin{eqnarray}
g_{ab^*}(F^a+g^{ac^*}\partial_{c^*} W^*)
(F^{*b}+g^{db^*}\partial_{d}W)
-g^{ab^*}\partial_{a}W \partial_{b^*} W^* ,
\end{eqnarray}
one finds that the action
\begin{eqnarray}
\parity: 
F^a+g^{ac^*}\partial_{c^*} W^*
\to
F^{*b}+g^{db^*}\partial_{d}W
\label{SU(2):F}
\end{eqnarray}
is a symmetry.
Secondly, 
we consider
the $\psi\psi$ and $\lambda\lambda$
mass terms in (\ref{L;K}), (\ref{L:superpotential})
and (\ref{L:gauge}).
Under the action (\ref{SU(2):fermion}) and (\ref{SU(2):F})
the $\psi\psi$ mass terms become
\begin{eqnarray}
\left(
\frac{i}{4}\CF_{abc}(F^c+g^{cd^*}\partial_{d^*} W^*)
-\frac{i}{4}\CF_{abc}g^{dc^*}\partial_dW
-\frac{1}{2}\partial_a\partial_b W
\right)\lambda^a\lambda^b.
\end{eqnarray}
Equating it with the original $\lambda\lambda$ mass term,
$\frac{i}{4}\partial_c\tau_{ab}F^c\lambda^a\lambda^b$,
we find that the invariance implies
\begin{eqnarray}
\frac{i}{4}\CF_{abc}
(g^{cd^*}\partial_{d*} W^*-g^{dc^*}\partial_{d} W)
-\frac{1}{2}\partial_a\partial_b W=0.
\end{eqnarray}
It is easy to see that the superpotential (\ref{W})
solves this equation.
Thirdly,
we examine the $\psi\lambda$ mass term in (\ref{L;K}) 
and (\ref{L:gauge})
\begin{eqnarray}
\frac{1}{\sqrt{2}}(
g_{ac^*}k_b^*{}^{c}
+\frac{1}{2}\partial_a\tau_{bc}D^c)\psi^a\lambda^b.
\end{eqnarray}
We rewrite it as
\begin{eqnarray}
\frac{1}{\sqrt{2}}\left(
g_{ac^*}k_b^*{}^{c}
-\frac{1}{4}\partial_a\tau_{bc}g^{cd}\fD_d
\right)\psi^a\lambda^b
+\frac{1}{2\sqrt{2}}\partial_a\tau_{bc}(D^c+\frac{1}{2}g^{cd}
\fD_d
)
\psi^a\lambda^b.
\end{eqnarray}
The invariance of the first term is guaranteed by
(\ref{psi-lambda}),
and thus we find
\begin{eqnarray}
\parity: 
D^c+\frac{1}{2}g^{cd}\fD_d
\to
-( D^c+\frac{1}{2}g^{cd}\fD_d
)
\label{SU(2):D}
\end{eqnarray}
for the invariance.
Lastly,
let us turn to the 
bosonic terms including $D^a$
\begin{eqnarray}
\frac{1}{2}(\tau_2)_{ab}D^aD^b
+\frac{1}{2}D^a
\left(
\fD_a~+2\sqrt{2} \xi \delta_{a}^0
\right).
\label{bosonic terms including D}
\end{eqnarray}
We rewrite it as 
\begin{eqnarray}
&&
\frac{1}{2}g_{ab}
\left(D^a+\frac{1}{2}g^{ac}(\fD_c+2\sqrt{2}\xi\delta_{c}^0)\right)
\left(D^b+\frac{1}{2}g^{bd}(\fD_d+2\sqrt{2}\xi\delta_{d}^0)\right)
\nonumber\\&&~~~
-\frac{1}{8}g^{ab}
\left(
\fD_a +2\sqrt{2} \xi \delta_{a}^0
\right)
\left(
\fD_b +2\sqrt{2} \xi \delta_{b}^0
\right).
\label{D square}
\end{eqnarray}
The first term in (\ref{D square}) is obviously invariant under
the action (\ref{2.***}) and (\ref{SU(2):D}).
The last term is also invariant
under the action (\ref{2.***})
because the term linear in $\xi$ vanishes
as is shown in (\ref{xi linear}).

As a result, we have found that the off-shell action $\CL$
(\ref{action:off-shell})
is invariant under the discrete  $\parity$-symmetry
(\ref{SU(2):fermion}),
(\ref{2.***}),
 (\ref{SU(2):F}) and (\ref{SU(2):D})
if we choose $\tau_{ab}$ as (\ref{tau})
and $W$ as (\ref{W}).
For completeness, we present the off-shell action of our $U(N)$ gauge model
which is invariant under the discrete  $\parity$-symmetry;
\begin{eqnarray}
\CL&=&
-g_{ab^*}\CD_mA^a\CD^mA^{*b}
-\frac{1}{4}
g_{ab}v_{mn}^av^{bmn}
-\frac{1}{8}
{\rm Re}
(\CF_{ab})\epsilon^{mnpq}v_{mn}^av_{pq}^b
\nonumber\\&&
-\frac{1}{2}\CF_{ab}\lambda^a\sigma^m\CD_m\bar\lambda^b
-\frac{1}{2}\CF_{ab}^*\CD_m\lambda^a\sigma^m\bar\lambda^b
-\frac{1}{2}\CF_{ab}\psi^a\sigma^m\CD_m\bar\psi^b
-\frac{1}{2}\CF_{ab}^*\CD_m\psi^a\sigma^m\bar\psi^b
\nonumber\\&&
+g_{ab^*}F^aF^{*b}
+F^a\partial_a W
+F^{*a}\partial_{a^*}  W^*
+\frac{1}{2}
g_{ab}D^aD^b
+\frac{1}{2}D^a \left( \fD_a + 2\sqrt{2} \xi \delta_{a}^0 \right)
\nonumber\\&&
+(\frac{i}{4}\CF_{abc}F^{*c}-\frac{1}{2}\partial_a\partial_b W)
 \psi^a\psi^b
+\frac{i}{4}\CF_{abc}
F^c\lambda^a\lambda^b 
+\frac{1}{\sqrt{2}}(g_{ac^*}k_{b}^*{}^{c}+\frac{1}{2}\CF_{abc}D^c)
\psi^a\lambda^b
\nonumber\\&&
+(-\frac{i}{4}\CF^*_{abc}F^{c}-\frac{1}{2}\partial_{a^*}
\partial_{b^*}  W^*)
\bar\psi^a\bar\psi^b 
-\frac{i}{4}\CF^*_{abc}
F^{*c}\bar\lambda^a\bar\lambda^b
+\frac{1}{\sqrt{2}}(g_{ca^*}k_{b}{}^{c}+\frac{1}{2}\CF^*_{abc}D^c)
\bar\psi^a\bar\lambda^b
\nonumber\\&&
-i\frac{\sqrt{2}}{8}(
\CF_{abc}
\psi^c\sigma^n\bar\sigma^m\lambda^a
-\CF^*_{abc}
\bar\lambda^a\bar\sigma^m\sigma^n\bar\psi^c
)v_{mn}^b
\nonumber\\&&
-\frac{i}{8}\CF_{abcd}
\psi^c\psi^d\lambda^a\lambda^b
+\frac{i}{8}\CF^*_{abcd}
\bar\psi^c\bar\psi^d\bar\lambda^a\bar\lambda^b, \label{off-shellaction2}
\end{eqnarray}
where
$g_{ab^*}={\rm Im}(\CF_{ab})$ and $W=eA^0+m\CF_0$.
In the above expression, 
the covariant derivatives
are defined as
\begin{eqnarray}
\CD_m\Psi^a&=&
 \partial_m\Psi^a - \frac{1}{2}f^a_{bc}v_m^b\Psi^c,
~~~~\Psi^a=\{A^a, \psi^a, \lambda^a \},
\label{gauge covariant derivative}
\\
v_{mn}^a&=&
 \partial_mv_n^a-\partial_nv_m^a
 -\frac{1}{2}f^a_{bc}v_m^bv_n^c.
\end{eqnarray}
By the reasoning we explained at the beginning of this section, our action (\ref{action:on-shell}) and (\ref{off-shellaction2}) are 
invariant under $\mathcal{N}=2$ supersymmetry.

\section{Extended Supersymmetry Transformation}

Our action is manifestly invariant under the $\mathcal{N}=1$ supersymmetry transformation. 
We have made our action invariant under the discrete transformation $\parity$, 
and the algebra of extended supersymmetry permits us to argue for the invariance of our action under the 
extended $\mathcal{N}=2$ supersymmetry transformation. In this section, we will first lift the 
$\mathcal{N}=1$ supersymmetry transformation
\begin{eqnarray}
\delta_{\eta_1} A^a&=&
\sqrt{2}\eta_1\psi^a, \label{susytransofA}
\\
\delta_{\eta_1} \psi^a&=&
 i\sqrt{2}\sigma^m\bar\eta_1\CD_mA^a
 +\sqrt{2}\eta_1 F^a, 
\\
\delta_{\eta_1} v_m^a&=&
 i\eta_1\sigma_m\bar\lambda^a-i\lambda^a\sigma_m\bar\eta_1, \label{susytransofv}
\\
\delta_{\eta_1} \lambda^a&=&
\sigma^{mn}
\eta_1 v_{mn}^a+i\eta_1 D^a, \label{susytransoflambda}
\end{eqnarray}
to its $\mathcal{N}=2$ counterpart by exploiting the discrete symmetry $\parity$. We will subsequently examine 
$SU(2)$ covariance of the $\mathcal{N}=2$ supersymmetry transformation obtained.

Let us first form a following doublet of fermions;
\begin{eqnarray}
\boldsymbol{\lambda}_i^{\ a} &\equiv& 
\left(
\begin{array}{c}
\lambda^a \\
\psi^a
\end{array}
\right),
~~~~~~~~~~~~~
\boldsymbol{\lambda}^{ia} \equiv
 \epsilon ^{ij} \boldsymbol{\lambda}_j^{\ a} =
\left(
\begin{array}{c}
+\psi^a \\
-\lambda^a
\end{array}
\right), \\
\bar{\boldsymbol{\lambda}}^{ia} &\equiv&
 \Bar{\boldsymbol{\lambda}_i^{\ a}}= 
\left(
\begin{array}{c}
\bar{\lambda}^a \\
\bar{\psi}^a
\end{array}
\right),
~~~~
\bar{\boldsymbol{\lambda}}_i^{\ a} \equiv
 \epsilon_{ik} \bar{\boldsymbol{\lambda}}^{ka} = 
\left(
\begin{array}{c}
-\bar{\psi}^a \\
+\bar{\lambda}^a
\end{array}
\right)
=-\Bar{\boldsymbol{\lambda}^{ia}} .
\end{eqnarray}
We carry out the raising and the lowering of $i,j$ indices by $\epsilon^{ij}$ ; $\epsilon^{12}=\epsilon_{21}=1$, $\epsilon^{21}=\epsilon_{12}=-1$.
Recall the action of $\parity$;
\begin{eqnarray}
\parity :\boldsymbol{\lambda}_i^{\ a} =
\left(
\begin{array}{c}
+\lambda^a \\
+\psi^a
\end{array}
\right)
&\longrightarrow& 
\boldsymbol{\lambda}^{ia}=
\left(
\begin{array}{c}
+\psi^a \\
-\lambda^a
\end{array}
\right), \nonumber \\
\bar{\boldsymbol{\lambda}}_i^{\ a}=
\left(
\begin{array}{c}
-\bar{\psi}^a \\
+\bar{\lambda}^a
\end{array}
\right)
&\longrightarrow&
\bar{\boldsymbol{\lambda}}^{ia}=
\left(
\begin{array}{c}
+\bar{\lambda}^a \\
+\bar{\psi}^a
\end{array}
\right),
\end{eqnarray}
and therefore the
terms $\hat F^a$ in (\ref{Fhat}) and $\hat D^a$
in (\ref{Dhat})  
which are bilinear in fermions undergo the action;
\begin{eqnarray}
\parity :
  \begin{array}{l}
\hat{F}^a \longrightarrow \hat{F}^{*a} ~       \\
\hat{D}^a \longrightarrow -\hat{D}^a ~       \\
  \end{array}
~.
\end{eqnarray}
Note that this is nothing but (\ref{SU(2):D}), (\ref{SU(2):F}).
The bosonic fields $A^a$, $v_m^{a}$ are invariant under $\parity$. 
So from (\ref{susytransofA}), (\ref{susytransofv}), we see that the grassman parameter $\eta_2$ for the second supersymmetry forms a doublet with $\eta_1$ such that 
\begin{eqnarray}
\parity :\boldsymbol{\eta}_i \equiv 
\left(
\begin{array}{c}
\eta_1 \\
\eta_2
\end{array}
\right)
\longrightarrow 
\left(
\begin{array}{c}
+\eta_2 \\
-\eta_1
\end{array}
\right)
\equiv \boldsymbol{\eta^i} \equiv \epsilon^{ij} \boldsymbol{\eta}_j \ \ .
\end{eqnarray}

Demanding the covariance under $\parity$, we obtain the extended supersymmetry transformation;
\begin{eqnarray}
\boldsymbol{\delta} A^a &=&
\sqrt{2} \boldsymbol{\eta}_j \boldsymbol{\lambda}^{ja}, \label{four} \\
\boldsymbol{\delta \lambda}_j^{\ a} &=&
\sigma^{mn} \boldsymbol{\eta}_j v_{mn}^{a} 
+ \sqrt{2}i \left( \sigma^m \bar{\boldsymbol{\eta}}_j \right) \mathcal{D}_m A^a
+ i
\left(
\begin{array}{cc}
\hat{D}^a            & +i\sqrt{2}\hat{F}^{*a} \\
-i\sqrt{2} \hat{F}^a & -\hat{D}^a
\end{array}
\right)
\left(
\begin{array}{c}
\eta_1 \\
\eta_2
\end{array}
\right) 
-\frac{i}{2} \boldsymbol{\eta}_j g^{ab} \fD_b
 \nonumber \\
& &
-\sqrt{2}ig^{ab^*} \frac{\partial}{\partial A^{*b}}
\left(
\begin{array}{cc}
\xi A^{*0}                       & +i(eA^{*0} + m{\mathcal{F}}^*_0) \\
-i(eA^{*0}+m{\mathcal{F}}^*_0) & -\xi A^{*0}
\end{array}
\right)
\left(
\begin{array}{c}
\eta_1 \\
\eta_2
\end{array}
\right),
\label{five} \\
\boldsymbol{\delta} v_m^{a} &=&
i \boldsymbol{\eta}_j \sigma_m \bar{\boldsymbol{\lambda}}^{ja}   
-i \boldsymbol{\lambda}_j^{\ a}{\sigma}_m\bar{\boldsymbol{\eta}}^j \ .
   \label{six}
\end{eqnarray}
Here
\begin{eqnarray}
\bar{\boldsymbol{\eta}}^j \equiv 
\left(
\begin{array}{c}
\bar{\eta}_1 \\
\bar{\eta}_2
\end{array}
\right)
\ \ \ 
\textrm{and}
\ \ \ 
\bar{\boldsymbol{\eta}}_j \equiv \epsilon_{ji} \bar{\boldsymbol{\eta}}^i
=
\left(
\begin{array}{c}
-\bar{\eta}_2 \\
+\bar{\eta}_1
\end{array}
\right). 
\label{seven}
\end{eqnarray}
The transformation (\ref{five}) is further recast into the following form;
\begin{eqnarray}
\boldsymbol{\delta \lambda}_j^{\ a} &=& 
(\sigma^{mn} \boldsymbol{\eta}_j)v_{mn}^{a}+\sqrt{2}i(\sigma^m \bar{\boldsymbol{\eta}}_j) 
\mathcal{D}_m A^a+i(\boldsymbol{\tau} \cdot \boldsymbol{D}^a)_j^{\ k} \boldsymbol{\eta}_k
-\frac{1}{2} \boldsymbol{\eta}_j f^a_{\ bc} A^{*b} A^c, \ \label{3.16} \\
\boldsymbol{\delta} \bar{\boldsymbol{\lambda}}^{ja} &=& -(\bar{\boldsymbol{\eta}}^j \bar{\sigma}^{mn})v_{mn}^{a} 
-\sqrt{2}i(-\boldsymbol{\eta}^j \sigma^m) \mathcal{D}_m A^{*a}-i\bar{\boldsymbol{\eta}}^k(\boldsymbol{\tau} \cdot \boldsymbol{D}^{*a})_k^{\ j} 
-\frac{1}{2} \bar{\boldsymbol{\eta}}^j f^a_{\ bc} A^b A^{*c}, \ \label{3.17} \\
\boldsymbol{D}^a &=&\hat {\boldsymbol{D}}^a -\sqrt{2} g^{ab^*} \frac{\partial}{\partial A^{*b}}
\left( \boldsymbol{\mathcal{E}}A^{*0}+\boldsymbol{\mathcal{M}}
{\mathcal{F}}_0^* \right). \label{3.18}
\end{eqnarray}
Here
\begin{eqnarray}
\hat{\boldsymbol{D}}^a&=&
(\hat{D}_1^a, \ \hat{D}_2^a, \ \hat{D}_3^a), 
~~~~\left\{
  \begin{array}{rcl}
\hat{D}_1^a+i\hat{D}_2^a       &=    &-i\sqrt{2} \hat{F}^a,    \\
\hat{D}_1^a-i\hat{D}_2^a       &=    &+i\sqrt{2} \hat{F}^{*a},    \\
\hat{D}_3^a       &=    &\hat{D}^a,    \\
  \end{array}
\right. \label{eight} 
\\
\boldsymbol{\mathcal{E}}&=&(0,\ -e,\ \xi), \label{nine} \\
\boldsymbol{\mathcal{M}}&=&(0,\ -m,\ 0), \label{ten}
\end{eqnarray}
and $\boldsymbol{\tau}$ are the Pauli matrices. We have used (\ref{deform:KP}) in the last term of (\ref{3.16}) and that of (\ref{3.17}).
Finally, we can easily check that (\ref{parityalgebra}) in fact holds in these transformations.

Let us now examine the $SU(2)$ covariance of the extended susy transformation given by (\ref{four}), (\ref{five}), (\ref{six}). 
All except the last term in (\ref{five}) are manifestly covariant under the rigid $SU(2)$ transformations. 
In particular, $\hat{\boldsymbol{D}}^a$, given by (\ref{Dhat})$\sim $(\ref{Fhat*}) which are bilinear in fermions, transforms as a real triplet under $SU(2)$,
\begin{eqnarray}
i \boldsymbol{\tau} \cdot \hat{\boldsymbol{D}}^a=
i\sqrt{2}
\left(
\begin{array}{cc}
\hat{D}^a & i\sqrt{2} \hat{F}^{*a} \\
-i\sqrt{2} \hat{F}^a & -\hat{D}^a
\end{array}
\right)=g^{ab^*}g_{cb^*,d} \boldsymbol{\lambda}_j^{\{ c} \boldsymbol{\lambda}^{d\} k}+g^{ab^*}g_{cb^*,d^*} \bar{\boldsymbol{\lambda}}_j^{\{ c} \bar{\boldsymbol{\lambda}}^{d\} k}.
\end{eqnarray}

The last term in (\ref{five}) is 
$SU(2)$ covariant provided the two three-dimensional real vectors $\boldsymbol{\mathcal{E}}$ and $\boldsymbol{\mathcal{M}}$ 
transform as triplets. 
Their actual form (\ref{nine}) and (\ref{ten}) tell us that
the rigid $SU(2)$ has been gauge fixed in this six-dimensional parameter space of ($\boldsymbol{\mathcal{E}}$, $\boldsymbol{\mathcal{M}}$), 
by making these two vectors point to a specific direction. 
The manifest $SU(2)$ covariance is lost at this point.
The transformation law we have exhibited generalizes 
the one seen in the literature \cite{Sohnius} by the inclusion of the $\xi$ term and the superpotential.

A very important property of the triplet of the auxiliary fields
$\boldsymbol{D}^a$ is that it is complex as opposed to be real.
Indeed, it has a constant imaginary part;
\begin{eqnarray}
\textrm{Im } \boldsymbol{D}^a = \delta^a_{\ 0} (-\sqrt{2} m)
\left(
0,
1,
0
\right). \label{2.**}
\end{eqnarray}
This supplies an essential ingredient for partial breaking of $\mathcal{N}=2$ supersymmetry in the next section.

The supersymmetry transformation law for the auxiliary fields is
determined by requiring the
closure of the $\eta_1$- and $\eta_2$-supersymmetries;
\begin{eqnarray}
\delta F^a&=&
-i\sqrt{2}\CD_m\psi^a\sigma^m\bar\eta_1
-\bar\eta_1\bar\lambda^b k_b{}^a
\nonumber\\&&
+\delta_{\eta_2}(g^{ab}\partial_bW-g^{ab}\partial_{b^*} W^*)
+i\sqrt{2}\eta_2\sigma^m\CD_m\bar\lambda^a
+\eta_2\psi^b k_b^*{}^a,\\
\delta D^a&=&
-\eta_1\sigma^m\CD_m\bar\lambda^a
-\CD_m\lambda^a\sigma^m\bar\eta_1
\nonumber\\&&
-\delta_{\eta_2}(g^{ab}\fD_b)
-\eta_2\sigma^m\CD_m\bar\psi^a
-\CD_m\psi^a\sigma^m\bar\eta_2,
\end{eqnarray}
where the $\CD_m$ represents the gauge covariant derivative
(\ref{gauge covariant derivative}).
The supersymmetry transformation forms the algebra
\begin{eqnarray}
[\delta_{\eta},\delta_{\eta'}]\Psi^a&=&
-2i(\eta\sigma^m\bar\eta'-\eta'\sigma^m\bar\eta)\CD_m\Psi^a,
~~~
\Psi^a=\{A^a,\psi^a,F^a,v_{mn}^a,\lambda^a,D^a\}
\end{eqnarray}
where
$(\eta,\eta')=(\eta_1,\eta'_1)$ or $(\eta_2,\eta'_2)$.

\section{Some Properties of the vacuum}
In order to discuss properties of our model, let us
fix the form of $\CF$.
The first equation in (\ref{adjoint})
implies that $k_a{}^b=f^b_{ac}A^c$
and thus $k_0{}^a=k_a{}^0=0$,
while the second equation in (\ref{adjoint}) implies that
\begin{eqnarray}
k_{\hat a}{}^{\hat b}\partial_{\hat b}\CF_0=0,
\label{CF:condition}
\end{eqnarray}
as well as 
$k_{\hat a}{}^{\hat b}\partial_{\hat b}\CF_{\hat c}=
-f^{\hat b}_{\hat a\hat c}\CF_{\hat b}$.
An obvious solution to (\ref{CF:condition})
is
\begin{eqnarray}
\CF=f(A^0)
+cA^0~\CG(\hat{B}) 
+\hat\CF(\hat A),
\label{prepotential:solution}
\end{eqnarray}
where $f(A^0)$, $\CG(\hat{B})$ and 
$\hat\CF(\hat A)$ are
analytic functions of $A^0$, $\hat{B}=\Tr(\hat A^2)/2c_2$ 
and a trace function of
$\hat A=A^{\hat a}t_{\hat a}$, respectively.
We can choose $\CG(0)=0$ without loss of generality. 
The constant $c_2$
is the quadratic Casimir 
defined by
 $\Tr(t_{\hat a}t_{\hat b})=c_2\delta_{\hat a\hat b}$.
One finds that for this prepotential the K\"ahler metric becomes
\begin{eqnarray}
g_{00^*}={\rm Im}(f_{00}),~~~
g_{\hat a0^*}=\delta_{\hat a\hat b}{\rm Im}(\CG'cA^{\hat b}),~~~
g_{\hat a\hat b^*}=
{\rm Im}\left(cA^0(
\delta_{\hat a\hat b}\CG'+\CG''\delta_{\hat a\hat c}\delta_{\hat b\hat d}
A^{\hat c}A^{\hat d})
+\hat\CF_{\hat a\hat b}
\right).
\end{eqnarray}
Note that the $U(1)$ part and the $SU(N)$ part have non-trivial mixings
as long as $c\neq 0$.
In the following we examine the model specified by
(\ref{prepotential:solution}).

Let us first examine
the local minimum
of the scalar  potential $\CV\equiv-\CL_{\textrm{pot}}$
\begin{eqnarray}
\CV
&=& g^{ab}\left(
\frac{1}{8}\fD_a\fD_b
+\xi^2\delta_{a}^0\delta_{b}^0
+\partial_a W\partial_{b^*} W^*
\right) \nonumber \\
&=& g^{ab} \left( \frac{1}{8} \fD_a \fD_b +
\partial_a \left( \boldsymbol{\mathcal{E}} A^0 
 + \boldsymbol{\mathcal{M}} 
\mathcal{F}_0 \right)
\cdot
\partial_{b^*} 
\left( \boldsymbol{\mathcal{E}} A^0 
 + \boldsymbol{\mathcal{M}} \mathcal{F}_0 \right)^* \right),
\end{eqnarray}
where we have used (\ref{xi linear}).
Here, we consider the unbroken $SU(N)$  phase
at which the $A^{\hat a}$ do not acquire vacuum expectation values.
Substituting $A^{\hat a}=0$ into 
the equation
\begin{eqnarray}
0=\delta\CV/\delta A^a&=&
-g^{bd}\partial_ag_{de}g^{ec}\left(
\frac{1}{8}\fD_b\fD_c
+\xi^2\delta_{b}^0\delta_{c}^0
+\partial_b W\partial_{c^*} W^*
\right)
\nonumber\\&&
+g^{bc}\left(
\frac{1}{4}\fD_b \partial_a\fD_c
+\partial_a \partial_b W\partial_{c^*} W^*
\right)~,\label{minimumofpotential}
\end{eqnarray}
we obtain
\begin{eqnarray}
\frac{i}{2}f_{000}g_{00}^{-2}\delta_{a}^0
\left(\xi^2+(e+mf_{00})(e+m f^*_{00})\right)
+g_{00}^{-1}
 mf_{000}\delta_{a}^0(e+m f^*_{00})=0.
 \label{minimum:equation}
\end{eqnarray}
Here we have derived
\begin{eqnarray}
\left\langle \fD_a \right\rangle =0,~~
\left\langle  \partial_aW \right\rangle =\delta_{a}^0(e+mf_{00}),~~
\partial_0\partial_a W 
=\delta_{a}^0mf_{000},~~
\partial_ag_{00}=-\frac{i}{2}f_{000}\delta_{a}^0,
\end{eqnarray}
as well as
\begin{eqnarray}
\left<g^{00}\right>=g_{00}^{-1},~~~
\left< g^{0\hat a} \right>=0. \label{5.8}
\end{eqnarray}
The expressions with bracket $\left\langle \cdots \right\rangle$
imply
$\cdots$ evaluated at $A^{\hat a}=0$.
It is obvious that (\ref{minimum:equation})
is satisfied when $f_{000}=0$,
but it is a saddle point because
$\left<\partial_0\partial_{0^*}\CV\right>=0$
, and thus does not represent a stable vacuum.
The stable minimum is at
\begin{eqnarray}
f_{00}=
 -\frac{e}{m}
\pm i\frac{\xi}{m}.
\label{minimum}
\end{eqnarray}

We shall show that at the stable minimum (\ref{minimum}) 
massless fermions emerge.
For this purpose, we examine the fermion mass term
\begin{eqnarray}
\CL_{\rm{mass}}&=&
-\frac{i}{4}g^{cd^*}\partial_c\tau_{ab}\partial_{d^*} W^*
 (\psi^a\psi^b+\lambda^a\lambda^b)
\nonumber\\&&
+\frac{1}{2\sqrt{2}}\left(
 g_{ac^*}k_b^*{}^{c}-g_{bc^*}k_a^*{}^{c}
 -\sqrt{2}\xi\delta_{c}^0(\tau^{-1}_2)^{cd}\partial_a\tau_{bd}
 \right) \psi^a\lambda^b~~+c.c. \ \ .
\end{eqnarray}
Substituting $A^{\hat a}=0$ into this mass term $\CL_{\rm mass}$,
we find that the the $U(1)$ fermions and the $SU(N)$ fermions
decouple because $\left\langle \mathcal{F}_{00\hat{a}} \right\rangle=0$,
\begin{eqnarray}
\CL_{{\rm mass}}&=&
\frac{1}{2}\boldsymbol{\lambda}_i^{\ 0} M_{U(1)}^{ij} \boldsymbol{\lambda}_j^{\ 0}
+\frac{1}{2}\delta_{\hat a\hat b} \boldsymbol{\lambda}_i^{\ \hat{a}} M_{SU(N)}^{ij} \boldsymbol{\lambda}_j^{\ \hat{b}}
~~+c.c.,
\nonumber\\
M_{U(1)}^{ij}&=&
-\frac{i}{2}g_{00}^{-1}f_{000}\left(
  \begin{array}{cc}
  e+m f^*_{00}     &-i\xi    \\
  -i\xi     &e+m f^*_{00}    \\
  \end{array}
\right),
\nonumber\\
M_{SU(N)}^{ij}&=&
-\frac{i}{2}g_{00}^{-1}c \left\langle \CG' \right\rangle
\left(
  \begin{array}{cc}
  e+m f^*_{00}     &-i\xi    \\
  -i\xi     &e+m f^*_{00}    \\
  \end{array}
\right).
\end{eqnarray}
It is easy to diagonalize these mass matrices
and one finds that
the $U(1)$ fermions $\frac{1}{\sqrt{2}}(\lambda^0\pm\psi^0)$
acquire masses $ \left| -\frac{i}{2}g_{00}^{-1}f_{000}(e+m f^*_{00}\mp i\xi) \right|$,
while the $SU(N)$ fermions 
$\frac{1}{\sqrt{2}}(\lambda^{\hat a}\pm\psi^{\hat a})$
acquire masses 
$\left| -\frac{i}{2}g_{00}^{-1}c\left\langle \CG'\right\rangle 
(e+m f^*_{00}\mp i\xi) \right|$.

At the stable minimum $f_{00}=-\frac{e}{m}\pm i\frac{\xi}{m}$,
the $U(1)$ fermion $\frac{1}{\sqrt{2}}(\lambda^0\mp\psi^0)$
and the $SU(N)$ fermions
$\frac{1}{\sqrt{2}}(\lambda^{\hat a}\mp\psi^{\hat a})$
remain massless,
while  the $U(1)$ fermion $\frac{1}{\sqrt{2}}(\lambda^0\pm\psi^0)$
and the $SU(N)$ fermions 
$\frac{1}{\sqrt{2}}(\lambda^{\hat a}\pm\psi^{\hat a})$
become massive with masses, $\left| -m\llangle f_{000} \rrangle \right|$
and $\left| -mc \llangle \CG' \rrangle\right|$, respectively.
Here, $\llangle \cdots \rrangle$ is the expectation value of $\cdots$ at the vacuum.
The $U(1)$ massless fermion is regarded as the Nambu-Goldstone fermion.

Let us demonstrate this last statement from the transformation law
 (\ref{five}).
Taking the expectation value,
 we see
\begin{eqnarray}
\Llangle
\boldsymbol{\delta} \boldsymbol{\lambda}^0 
\Rrangle
&=& -\sqrt{2}i \Llangle g^{00} \Rrangle
\left(
\begin{array}{cc}
\xi & i\llangle e+mf^*_{00}\rrangle \\
-i\llangle e+mf^*_{00}\rrangle & -\xi
\end{array}
\right)
\left(
\begin{array}{c}
\eta_1 \\
\eta_2
\end{array}
\right) \nonumber \\
&=&
\mp \sqrt{2} im
\left(
\begin{array}{cc}
1 & \pm 1 \\
\mp 1 & -1
\end{array}
\right)
\left(
\begin{array}{c}
\eta_1 \\
\eta_2
\end{array}
\right), \label{vacuum1} \\
\Llangle
 \boldsymbol{\delta} \boldsymbol{\lambda}^{\hat{a}} 
\Rrangle &=& 0. \label{vacuum2}
\end{eqnarray}
We have used (\ref{5.8}), (\ref{minimum}).
Therefore,
\begin{eqnarray}
\LLangle
 \frac{\delta(\lambda^0 \mp \psi^0)}{\sqrt{2}} 
\RRangle &=& \mp 2mi(\eta_1 \pm \eta_2)
, \nonumber \\
\LLangle
 \frac{\delta(\lambda^0 \pm \psi^0)}{\sqrt{2}}
\RRangle &=& 0. 
\end{eqnarray}
One linear combination of the $U(1)$ fermion,
$\frac{1}{\sqrt{2}}(\lambda^0\mp \psi^0)$,
 is in fact the Nambu-Goldstone fermion.

Finally, let us discuss a mechanism which is responsible for partial breaking of
 $\mathcal{N}=2$ supersymmetry to be realized. 
We see that partial breaking requires that the $2 \times 2$ matrix 
$\llangle \boldsymbol{\tau} \cdot \boldsymbol{D}^a \rrangle$
in (\ref{3.16})
 has one nonvanishing eigenvalue for some $a$.
We obtain
\begin{eqnarray}
-\llangle \det \boldsymbol{\tau} \cdot \boldsymbol{D}^a \rrangle &=&
 \llangle \boldsymbol{D}^a \cdot \boldsymbol{D}^a \rrangle \nonumber \\
&=& \llangle \Re \boldsymbol{D}^a \cdot \Re \boldsymbol{D}^a \rrangle
-\llangle \Im \boldsymbol{D}^a \cdot \Im \boldsymbol{D}^a \rrangle
+2i \llangle \Re \boldsymbol{D}^a \cdot \Im \boldsymbol{D}^a \rrangle
\nonumber \\
&=& 0,
\end{eqnarray}
which implies that partial breaking is certainly not possible without nonvanishing imaginary part of $\boldsymbol{D}^a$.
Using (\ref{2.**}), we convert this condition into
\begin{eqnarray}
\Llangle \Re \boldsymbol{D}^{\hat a} \Rrangle&=&0, \nonumber \\
\Arrowvert \Llangle \Re \boldsymbol{D}^0 \Rrangle \Arrowvert =
  \Arrowvert \Im \boldsymbol{D}^0 \Arrowvert &=& 
\sqrt{2} m, \label{***} \\
\Llangle \Re \boldsymbol{D}^0 \Rrangle
 \cdot \Im \boldsymbol{D}^0 &=&0. \nonumber
\end{eqnarray}
Coming back to the extremum condition (\ref{minimumofpotential}) 
of the scalar potential at the unbroken $SU(N)$ phase, 
we see that it can also be converted as 
\begin{eqnarray}
0=\frac{\delta \mathcal{V}}{\delta A^a}=
 \frac{i}{4} \llangle f_{000} \delta_{a}^0 \rrangle 
 \Llangle \boldsymbol{D}^0 \Rrangle \cdot
  \Llangle \boldsymbol{D}^0 \Rrangle.
\end{eqnarray}
The condition for a stable vacuum is obviously equivalent to that of partial supersymmetry breaking (\ref{***}). 
Note that at the vacuum
\begin{eqnarray}
\llangle \mathcal{V} \rrangle = 
\Llangle g_{00}^{-1} \Rrangle 
\left( \xi^2+ \Llangle \left| e+mf_{00} \right|^2 \Rrangle \right)
=\pm 2m\xi 
\neq 0 ~.
\label{4.**}
\end{eqnarray}


\section{$\CN=2$ Supercurrents}

In the previous section, the rigid  $SU(2)$ symmetry, in particular,
 its discrete element $\parity$ has been exploited to provide $N=2$ supersymmetry
of our model. In this section, we discuss another rigid transformation, 
namely,  the one associated with the 
$U(1)_{R}$ transformation and the attendant supermultiplet of currents.

It is well known that the Wess-Zumino model consisting of
 the scalar superfield with a superpotential permits  
the $U(1)_R$ current,  the supercurrent  and the energy momentum tensor
 as a supermultiplet of currents when the superpotential is a monomial
 in scalar superfield \cite{FZ}. It is then possible to assign $R$ weight one 
to the superpotential.  
(Extended) supermultiplet of currents exists for ($\mathcal{N}=2$) super Yang-Mills as well \cite{FZ} \cite{IKT}. 
Starting from the $U(1)_R$ current, we can use this multiplet structure to derive the form 
of the supercurrent and the energy momentum tensor and to check the consistency of supersymmetry algebra. We illustrate
this in the Wess-Zumino model in the Appendix A. 
Our model has  $N=2$ supermultiplet of Noether currents when it is possible to assign $R$ weight two to the prepotential $\mathcal{F}$. 
We show how this is used to derive the $\mathcal{N}=2$ supercurrents for generic $\mathcal{F}$.

The $R$ transformation is given by
\begin{eqnarray}
R:\ \ 
\Phi(x,\theta,\bar{\theta}) &\rightarrow& e^{i \alpha}
\Phi(x,e^{-\frac{i \alpha}{2}}\theta,\bar{\theta}), \nonumber \\
\mathcal{W}_{\alpha}(x,\theta,\bar{\theta}) &\rightarrow&
\mathcal{W} _{\alpha} (x,e^{-\frac{i\alpha}{2}}\theta,\bar{\theta}),
\end{eqnarray}
\begin{eqnarray}
R: A &\rightarrow& e^{i\alpha} A \ \ ,\ \ v_m \rightarrow v_m, \nonumber \\
\psi &\rightarrow& e^{\frac{i\alpha}{2}} \psi \ \ \ ,\ \ \ \lambda
\rightarrow e^{\frac{i\alpha}{2}\lambda}, \label{5.2} \\
F &\rightarrow& F \ \ \ \ \ ,\ \ \ \ \ D \rightarrow D. \nonumber
\end{eqnarray}
We assume that the prepotential $\mathcal{F}$ is transformed as weight two under $R$
\begin{equation}
\mathcal{F} \rightarrow e^{2i\alpha} \mathcal{F}.
\end{equation}
The $U(1)_R$ current associated is 
\begin{eqnarray}
\theta J \bar{\theta} &\equiv& (\tau_2)_{ab} \left( \bar{\theta}
\bar{\boldsymbol{\lambda}}^{ia} \boldsymbol{\lambda}_i^{\ b} \theta + iA^{*^a}\theta
\overleftrightarrow{\mathcal{D}} \cdot \sigma \bar{\theta} A^b \right)
\label{U(1)Rcurrent} \\
&\equiv& (\tau_2)_{ab} \left( \theta j^{ab} \bar{\theta} +2\theta \Delta
j^{ab} \bar{\theta} \right) ~.
\end{eqnarray}
The second term is known as the improvement term.
Using the transformation law of rigid $\mathcal{N}=2$ supersymmetry in section III, we obtain
\begin{eqnarray}
\theta \boldsymbol{\delta} J \bar{\theta} = 
(\tau_2)_{ab} \left( \theta \boldsymbol{\delta} j^{ab} \bar{\theta} + 2\theta \boldsymbol{\delta} (\Delta j)^{ab} \bar{\theta} \right) 
+\boldsymbol{\delta} (\tau_2)_{ab} \left( \theta j^{ab} \bar{\theta} +2\theta (\Delta j)^{ab} \bar{\theta} \right),
\label{6.11}
\end{eqnarray}
where
\begin{eqnarray}
&&\!\!\!\!\!
\theta \boldsymbol{\delta} j^{ab} \bar{\theta} \nonumber\\
&&=\bar{\theta} \bar{\boldsymbol{\lambda}}^{jb}
\left( 
(\theta \sigma^{mn} \boldsymbol{\eta}_j)v_{mn}^{\ \ a} +\sqrt{2}i (\theta \sigma^m \bar{\boldsymbol{\eta}}_j) \mathcal{D}_m A^a 
+i(\boldsymbol{\tau} \cdot \boldsymbol{D}^a)_j^{\ k}(\theta \boldsymbol{\eta}_k) -\frac{1}{2}(\theta \boldsymbol{\eta}_j)f^a_{\ cd}A^{*c}A^d 
\right) \nonumber \\&&
-\left(
 (\bar{\boldsymbol{\eta}}^j \bar{\sigma}^{mn} \bar{\theta}) v_{mn}^{\ \ b}
 +\sqrt{2}i(-\boldsymbol{\eta}^j \sigma^m \bar{\theta}) \mathcal{D}_m A^{*b}
 +i(\bar{\boldsymbol{\eta}}^k \bar{\theta})(\boldsymbol{\tau} \cdot
  \boldsymbol{D}^{*b})_k^{\ j}
 +\frac{1}{2}(\bar{\boldsymbol{\eta}}^j \bar{\theta})f^b_{\ ef}A^e A^{*f}
\right)
\theta \lambda_j^{\ a}, \nonumber \\&&
\\
&&\!\!\!\!\!
\theta \boldsymbol{\delta} (\Delta j^{ab}) \bar{\theta} 
=\frac{\sqrt{2}}{2} i A^{*a} \theta \overleftrightarrow{\mathcal{D}}_m \sigma^m \bar{\theta} \boldsymbol{\eta}_j \boldsymbol{\lambda}^{jb}
+\frac{\sqrt{2}}{2}i \bar{\boldsymbol{\eta}}^j \bar{\boldsymbol{\lambda}}_j^{\ a} \theta \overleftrightarrow{\mathcal{D}}_m \sigma^m \bar{\theta} A^b
+
\frac{i}{2} A^{*a} \theta \boldsymbol{\delta} 
\overleftrightarrow{\mathcal{D}}_m \sigma^m \theta A^b ,
\label{improved}\\
&&\!\!\!\!\!
2i \boldsymbol{\delta} (\tau_2)_{ab} = 
\sqrt{2} 
\left( 
\tau_{abc}(A^d) \boldsymbol{\eta}_i \boldsymbol{\lambda}^{ic} -\tau^*_{abc} (A^{*d}) \bar{\boldsymbol{\eta}}^i \bar{\boldsymbol{\lambda}}_i^{\ c}
\right).
\end{eqnarray}
In the case where the prepotential is a degree two polynomial in $A^a$, $\boldsymbol{\delta} (\tau_2)_{ab}=0$ and eq.(\ref{6.11}) provides 
construction of $\mathcal{N}=2$ improved supercurrents which are conserved;
\begin{eqnarray}
\boldsymbol{\eta}_j \boldsymbol{\mathcal{S}}^{(j)m}+\bar{\boldsymbol{\eta}}^j \bar{\boldsymbol{S}}_{(j)}^{\ m} \equiv
-\frac{1}{2} (\tau_2)_{ab} \tr \bar{\sigma}^m \left( \boldsymbol{\delta} (j^{ab})+2\boldsymbol{\delta} (\Delta j^{ab}) \right) .
\end{eqnarray}
Here `` $\tr$ " implies a trace in the spinor space.

The $R$ current is not conserved when $\mathcal{F}$ is not a degree two polynomial in $A$ and the above construction would appear not useful for the general construction 
of the conserved supercurrents. We will show below that this is not the case. Let us write the prepotential $\mathcal{F}$ generically as 
\begin{eqnarray}
\mathcal{F}= \sum_{n,j} h_j^{(n)}C_j^{(n)}(A^a). \label{*}
\end{eqnarray}
Here $C_j^{(n)}(A^a)$ are $n$-th order $U(N)$ invariant polynomials in $A^a$ properly normalized and labelled by the index $j$, and $h_j^{(n)}$ are their coefficients. 
We first observe that we can assign weight two to $\mathcal{F}$ in (\ref{*}) provided $h_j^{(n)}$ transform as weight $-(n-2)$. 
Let us consider the local version of the $U(1)_R$ transformation (\ref{5.2}), replacing $\alpha$ by $\alpha (x)$. We obtain
\begin{eqnarray}
&&S[A e^{i\alpha(x)},\ \boldsymbol{\lambda}_j e^{\frac{i\alpha (x)}{2}},\ ...] 
-S [A,\ \boldsymbol{\lambda}_j,\ ...] 
\nonumber \\&&
\ \ =\int d^4 x\  \partial_m \left( \alpha (x) \left( -\frac{1}{2} \right)
 \tr \bar{\sigma}^m J \right)
+\int d^4 x\  \alpha (x) \partial_m 
 \left( \frac{1}{2} \tr \bar{\sigma}^m J \right)
\nonumber\\&&
\ \ \ \ \ +\int d^4 x\ i \alpha (x) \sum_{n,j} (n-2) \frac{\partial}{\partial h_j^{(n)}} \mathcal{L}.
\end{eqnarray}
Here $\mathcal{L}$ and $S$ are the Lagrangian and the action of our model respectively. The left hand side vanishes by the equation of motion, and we obtain
\begin{eqnarray}
\partial_m \left( -\frac{1}{2} \tr \bar{\sigma}^m J \right)
=i \left( \sum_{n,j} (n-2) \frac{\partial}{\partial h_j^{(n)}} \right) \mathcal{L} \equiv \Delta_h \mathcal{L}.
\end{eqnarray}
Taking the supersymmetry variation of this equation, we obtain
\begin{eqnarray}
\partial_m \left( -\frac{1}{2} \tr \bar{\sigma}^m \boldsymbol{\delta} J \right)=  \Delta_h \boldsymbol{\delta} \mathcal{L}.
\end{eqnarray}
As our action is $\mathcal{N}=2$ supersymmetric, the right hand side is written as 
\begin{eqnarray}
\Delta_h \partial_m X^m = \partial_m \Delta_h X^m , \\
X^m = \boldsymbol{\eta}_j \boldsymbol{y}^j + \bar{\boldsymbol{\eta}}^j \bar{\boldsymbol{y}}_j,
\end{eqnarray}
for some operator $X^m$ linear in $\boldsymbol{\eta}_i$ and $\bar{\boldsymbol{\eta}}^i$.
Hence
\begin{eqnarray}
\partial_m \left( -\frac{1}{2} \tr \bar{\sigma}^m \boldsymbol{\delta} J-\Delta_h X^m \right) = 0.
\end{eqnarray}
This provides a general construction of the conserved 
$\mathcal{N}=2$ supercurrents of our model;
\begin{eqnarray}
\boldsymbol{\eta}_j \boldsymbol{\mathcal{S}}^{(j)m}+\bar{\boldsymbol{\eta}}^j \bar{\boldsymbol{\mathcal{S}}}_{(j)}^{\ m} \equiv 
-\frac{1}{2} \tr (\bar{\sigma}^m \boldsymbol{\delta}J)-\Delta_h X^m . \label{transofcurrent}
\end{eqnarray}

The form of the supercurrents given 
in eq.(\ref{transofcurrent}) tells us 
that our model does not permit a universal coupling to $\mathcal{N}=2$
supergravity. 
The piece $-\Delta_h X^m$ is not generic 
and depends on the functional form of the prepotential
$\mathcal{F}(A)$ in $A$. 
This and the previous analysis in \cite{APT, central charge}
support the point of view  
that $\mathcal{N}=2$ supersymmetric gauge models 
with nontrivial K\"ahler potential should be viewed 
as a low energy effective action.

Let us now further transform (\ref{transofcurrent})
\begin{eqnarray}
\boldsymbol{\delta} \left( \boldsymbol{\eta}_j \boldsymbol{\mathcal{S}}^{(j)m} + \bar{\boldsymbol{\eta}}^j \bar{\boldsymbol{\mathcal{S}}}_{(j)}^{\ \ m} \right)
=-\frac{1}{2} \tr \bar{\sigma}^m \boldsymbol{\delta} \boldsymbol{\delta} J - \Delta_h \boldsymbol{\delta} X^m. \label{transofcurrent2}
\end{eqnarray}
This generates the $\mathcal{N}=2$ supersymmetry algebra (\ref{intro}) quoted in the introduction and at the same time provides its consistency conditions. Let us note that
\begin{eqnarray}
\theta \boldsymbol{\delta} \boldsymbol{\delta} J \bar{\theta} &=& 
(\tau_2)_{ab} \left( \theta \boldsymbol{\delta}
\boldsymbol{\delta} j^{ab} \bar{\theta} +2\theta \boldsymbol{\delta} \boldsymbol{\delta}(\Delta j)^{ab}
\bar{\theta} \right)
+2\boldsymbol{\delta} (\tau_2)_{ab}
  \left( \theta \boldsymbol{\delta} j^{ab} \bar{\theta}+2\theta
\boldsymbol{\delta}(\Delta j)^{ab} \bar{\theta} \right) \nonumber \\
& & +\boldsymbol{\delta} \boldsymbol{\delta} (\tau_2)_{ab}
  \left( \theta j^{ab} \bar{\theta}
+2\theta(\Delta j)^{ab} \bar{\theta} \right) \label{deltadeltaj}.
\end{eqnarray}
Denote by $\delta_{\eta_j}$ ($\delta_{\bar{\eta}^j}$) 
the transformation in which only $\eta_j$($\bar{\eta}^j$) 
is kept in $\boldsymbol{\delta}$.
The conditions
\begin{eqnarray}
\begin{array}{c}
\delta_{\eta_j} \mathcal{S}^{(j)m}=0 \\
\delta_{\bar{\eta}^j} \bar{\mathcal{S}}^{(j)m}=0 
\end{array}
\ \ \ \ \ \ \ \textrm{with $j$ not summed}
\end{eqnarray}
provide
\begin{eqnarray}
-\frac{1}{2} \tr \bar{\sigma}^m \delta_{\eta_j} \delta_{\eta_j} J-\Delta_h \eta_j \delta_{\eta_j} y^j =0 \ \ \ \ \ \ \ \textrm{with $j$ not summed} \label{**}
\end{eqnarray}
and its complex conjugate. Their actual expressions are quite involved as one sees from (\ref{deltadeltaj}) and the transformation laws (\ref{four})$\sim $(\ref{3.18}). We will not discuss eq.(\ref{**})
 further in this paper. In the case where $\mathcal{F}$ is degree two in $A$, $y^j=0$, and $\boldsymbol{\delta} (\tau_2)_{ab}=0$,
  eq.(\ref{**}) can be checked easily as in \cite{IKT} 
and in Appendix (\ref{app}) with the aid of the equations of motion. 

Let us finally read off the constant matrix $C_i^{\ j}$ in (\ref{intro}) from our algebra (\ref{transofcurrent2}). The only piece in (\ref{deltadeltaj}) which can contribute to 
$C_i^{\ j}$ is the part in $(\tau_2)_{ab} \boldsymbol{\delta} \boldsymbol{\delta} j^{ab}$ which is linear both in $\boldsymbol{D}^a$ and in $\boldsymbol{D}^{*a}$. 
This part is computed as 
\begin{eqnarray}
2 \left( \tau_2 \right)_{ab} \boldsymbol{D}^{*b} \cdot \boldsymbol{D}^a \bar{\theta} \left( \bar{\boldsymbol{\eta}} \boldsymbol{\eta} \right) \theta 
+2i\left( \tau_2 \right)_{ab} \left( \boldsymbol{D}^{*b} \times \boldsymbol{D}^a \right) \cdot \bar{\theta} \bar{\boldsymbol{\eta}} \boldsymbol{\tau} \boldsymbol{\eta} \theta .
\end{eqnarray}
Substituting the expressions (\ref{3.18}) $\sim $ (\ref{ten}) 
into this equation, we find that the second term contains $8m\xi \bar{\theta} \bar{\boldsymbol{\eta}} \tau_1 \boldsymbol{\eta} \theta$. Translated into (\ref{intro}), this implies
\begin{eqnarray}
C_i^{\ j}=+2m\xi (\boldsymbol{\tau}_1)_i^{\ j}.
\end{eqnarray}
This is consistent with (\ref{4.**}).

\section{Fermionic Shift Symmetry}

Equations (\ref{vacuum1}) and (\ref{vacuum2}) express the extended supersymmetry transformation of the $SU(2)$ doublet of $U(N)$ fermions on the vacuum as $U(1)$ fermionic 
shift generated by
\begin{eqnarray}
\boldsymbol{\chi}_i \equiv \sqrt{2} m
\left(
\begin{array}{c}
\eta_1 \\
\eta_2
\end{array}
\right)
=
\left(
\begin{array}{c}
\chi_1 \\
\chi_2
\end{array}
\right).
\end{eqnarray}
Note that the coupling constants $e$, $m$, $\xi$ of our model carry dimension two and that $\boldsymbol{\chi}_i$ carry dimension $3/2$. The Nambu-Goldstone fermion 
is the maximal mixing of the $U(1)$ gauge fermion and the $U(1)$ matter fermion.

Restricting our attention to the $U(N)$ field strength gauge superfield $\mathcal{W}_{\alpha}$, let us recast (\ref{vacuum1}) into
\begin{eqnarray}
\llangle
\boldsymbol{\delta} \mathcal{W}_{\alpha}
\rrangle
=
\left(
\mp \chi_1 - \chi_2
\right)
\boldsymbol{1}_{N\times N} \equiv
4\pi \chi_{\alpha} \boldsymbol{1}_{N\times N} .
\end{eqnarray}
We obtain
\begin{eqnarray}
\llangle \boldsymbol{\delta} S \rrangle &=&
\chi^{\alpha}  \llangle w_{\alpha} \rrangle, \\
\llangle \boldsymbol{\delta} w_{\alpha} \rrangle &=&
 N \chi_{\alpha},
\end{eqnarray}
where
\begin{eqnarray}
S=\frac{1}{32\pi^2} \tr \mathcal{W}^{\alpha} \mathcal{W}_{\alpha},\ \ \ \ w_{\alpha}=\frac{1}{4\pi} \tr \mathcal{W}_{\alpha}.
\end{eqnarray}
In this sense, our spontaneously broken supersymmetry is realized on the vacuum as the $U(1)$ fermionic shift noted by ref \cite{CDSW}
in the $\mathcal{N}=2$ $U(N)$ super Yang-Mills deformed by the superpotential $W(\Phi)$. 
See also \cite{IK1}. 
As for its transformation acting on the fields or equivalently on a generic state, let us note that 
\begin{eqnarray}
\boldsymbol{\delta} \boldsymbol{\lambda}_j^{\ a} =
 \Llangle \boldsymbol{\delta} \boldsymbol{\lambda}_j^{\ a} \Rrangle 
 + \cdots .
\end{eqnarray}
Here 
$\Llangle \boldsymbol{\delta} \boldsymbol{\lambda}_j^{\ a} \Rrangle$ 
is given in (\ref{vacuum1}), 
and $\cdots$ denotes the parts which do not receive the vacuum expectation values. 
This latter part is to be suppressed by $\frac{1}{m}$ with the replacement
$\boldsymbol{\eta}_j \rightarrow \frac{\boldsymbol{\chi}_j}{\sqrt{2} m}$ 
when
\begin{eqnarray}
\frac{e}{m} \ll 1,\ \ \ \frac{\xi}{m} \ll 1,\ \ \ \xi \neq 0
\end{eqnarray}
for appropriate low energy processes. The spontaneously broken supersymmetry operates as an approximate fermionic $U(1)$ shift symmetry in this regime.


\bigskip

\section*{Acknowledgements}

The authors thank 
Muneto Nitta, 
Kazutoshi Ohta, 
Norisuke Sakai 
and 
Yukinori Yasui for useful discussions.
This work is supported in part by the Grant-in-Aid for Scientific
Research(16540262) from the Ministry of Education,
Science and Culture, Japan.
Support from the 21 century COE program
``Constitution of wide-angle mathematical basis focused on knots"
is gratefully appreciated.
The preliminary version of this work was presented both in
SUSY2004, Tsukuba, Japan (June17-23 2004) and in YITP workshop YITP-W-04-03 on ``Quantum Field Theory 2004" in the Yukawa Institute
for Theoretical Physics, Kyoto University (July 13-16 2004).
We wish to acknowledge the participants
for stimulating discussions.
\newpage
\appendix

\section{A}

In the text, we exploited the extended operation $\parity$ which involves the sign change of the parameter $\xi$ as well as the transformation
of the two component spinor parameter $\boldsymbol{\eta}_j$ to demonstrate that our action $\mathcal{L}$ or $\mathcal{L}'$ is 
invariant under $\mathcal{N}=2$ supersymmetry. Though the use of $\parity$ is logical from an algebraic point of view, clearly it is not 
a symmetry in the sense of Noether. In this appendix, we provide another proof, using the more conventional operation  
which involves the transformation of the fields alone.
To be more specific, let $R$ be a generator such that
\begin{eqnarray}
R \lambda ^a R^{-1} = \psi ^a, \ R \psi ^a R^{-1}=-\lambda ^a, \ R A^a R^{-1} = A^a, \ \textrm{and} \ R v_m^{\ a} R^{-1}=v_m^{\ a}.
\label{A.1}
\end{eqnarray}

Let us start from the eqs. of the $\mathcal{N}=1$ transformation laws (\ref{susytransofA}) $\sim $ (\ref{susytransoflambda}), 
replacing $\eta$, by $\theta$ and writing $F^a$ and $D^a$ explicitly by (\ref{3.18}), (\ref{eight}).
\begin{eqnarray}
\delta_{\theta}^{(1,\xi)} A^a 
&=& \sqrt{2} \theta \psi^a, \label{A.2} \\
\delta_{\theta}^{(1,\xi)} \psi ^a 
&=& i \sqrt{2} \sigma^m \bar{\theta} \mathcal{D}_m A^a 
  + \sqrt{2} \theta \left( \hat{F}^a - \sqrt{2} g^{ab^*} \frac{\partial}{\partial A^{*b}} \left( e A^{*0} + m \mathcal{F}_0^* \right) \right), \\
\delta_{\theta}^{(1,\xi)} v_m^{\ a} 
&=& i \theta \sigma_m \bar{\lambda} ^a -i \lambda^a \sigma^m \bar{\theta}, \label{A.4} \\
\delta_{\theta}^{(1,\xi)} \lambda^a 
&=& \sigma^{mn} \theta v_{mn}^a 
 + i\theta \left( \hat{D}^a-\sqrt{2} g^{ab^*} \frac{\partial}{\partial A^{*b}} \left( \xi A^{*0} \right) \right), \label{A.5}
\end{eqnarray}
where $\hat{D}^a$ and $\hat{F}^a$ are given in terms of fermion bilinears by (\ref{Dhat}), (\ref{Fhat}). 
We have introduced the superscript $(1,\xi)$ to label the transformation fully. 
Operating $R$ from the left and $R^{-1}$ from the right on (\ref{A.5}), we obtain
\begin{eqnarray}
R \delta_{\theta}^{(1,\xi)} \lambda^a  R^{-1}
&=& \left( R \delta_{\theta}^{(1,\xi)} R^{-1} \right) \psi^a  \nonumber \\
&=& \sigma^{mn} \theta v_{mn}^{\ a} + i\theta \left( -\hat{D}^a-\sqrt{2} g^{ab^*} \frac{\partial}{\partial A^{*b}} \left( \xi A^{*0} \right) \right), \label{A.6}
\end{eqnarray}
where we have used $R \hat{D}^a R^{-1}=-\hat{D}^a.$ 
Eq. (\ref{A.6}) is compared with $\boldsymbol{\delta} \psi^a$ at $\eta_1=0$ in (\ref{five}) of the text, and we find
\begin{eqnarray}
R \delta_{\eta_1=\theta}^{(1,\xi)} R^{-1} = \boldsymbol{\delta}_{\eta_1=0,\eta_2=\theta}^{(-\xi)} 
\equiv \delta_{\eta_2=\theta}^{(2,-\xi)} \label{A.7}
\end{eqnarray}
on $\psi^a$. We have introduced the subscript and the superscript to $\boldsymbol{\delta}$ to specify the transformation
completely. Proceeding in a similar way on (\ref{A.1}), we obtain
\begin{eqnarray}
R \delta^{(1,\xi)} \psi^a R^{-1} 
&=& R \delta^{(1,\xi)} R^{-1} (-\lambda^a) \nonumber \\
&=& i\sqrt{2} \sigma^m \bar{\theta} \mathcal{D}_m A^a
   +\sqrt{2} \theta \left( \hat{F}^{*a}-\sqrt{2}g^{ab^*} \frac{\partial}{\partial A^{*b}} \left( eA^{*0}+m\mathcal{F}_0^* \right) \right). 
\end{eqnarray}
We see that (\ref{A.7}) is true on $\lambda^a$ as well.
It is easy to check from (\ref{A.2}), (\ref{A.4}) that (\ref{A.7}) holds on $A^a$ and on $v_m^a$. 
We conclude that (\ref{A.7}) is valid on all fields.

Once this is established, it is immediate to provide a proof that our action is invariant under $\mathcal{N}=2$ supersymmetry. 
Let
\begin{eqnarray}
S(\xi)=\int d^4 x \mathcal{L} (x) \ \ \textrm{or} \ \int d^4 x \mathcal{L'} (x),
\end{eqnarray}
where $\mathcal{L}(x)$ and $\mathcal{L}'(x)$ are given by (\ref{action:off-shell}) and by (\ref{action:on-shell}) respectively.
$\mathcal{N}=1$ supersymmetry implies
\begin{eqnarray}
\delta_{\eta_1=\theta}^{(1,\xi)} S(\xi)=0~.
\end{eqnarray}
Multiplying $R$ from left and $R^{-1}$ from right, we obtain
\begin{eqnarray}
\left( R \delta_{\eta_1=\theta}^{(1,\xi)} R^{-1} \right) R S(\xi) R^{-1} = \delta_{\eta_2=\theta}^{(2,-\xi)} S(-\xi) =0,\ \
\mbox{and thus} \ \ \
\delta_{\eta_2=\theta}^{(2,\xi)} S(\xi)=0,
\end{eqnarray}
which is a statement that our action is $\mathcal{N}=2$ supersymmetric. \\

\section{B}
In this appendix, we reexamine the current supermultiplet in the
Wess-Zumino model. While its superfield expression is well-known, 
we will present this supermultiplet in the component formalism, so that the
reasoning here is applicable to the discussion in the text.
The action is
\begin{equation}
S=\int d^4 x \mathcal{L} \ , \ \mathcal{L}=\int d^2 \theta d^2
\bar{\theta} {\Phi}^* \Phi+\int d^2 \theta W(\Phi)+\int d^2
\bar{\theta} {W}^*({\Phi}^*)
\end{equation}
and the superpotential $W(\Phi)$ $\left( \textrm{or} \
{W}^*({\Phi}^*) \right)$ is a monomial of degree $k$ in $\Phi$
$\left( \textrm{or} \ {\Phi}^* \right)$
. The model possesses $U(1)_R$ symmetry associated with
\begin{eqnarray}
&R:& \ \Phi(x,\theta,\bar{\theta}) \rightarrow e^{i\alpha /k}
\Phi(x,e^{-i\alpha /2}\theta,\bar{\theta}), 
\end{eqnarray}
so that
\begin{eqnarray}
&R:& \ A \rightarrow e^{i\alpha /k} A, \nonumber \\
& & \ \psi \rightarrow e^{i\alpha(\frac{1}{k}-\frac{1}{2})} \psi,
\label{Rsymmetry}\\
& & \ F \rightarrow e^{i\alpha(\frac{1}{k}-1)}F. \nonumber
\end{eqnarray}
The proper Noether current $J_{\alpha \dot{\alpha}}$ is given by
\begin{eqnarray}
\theta J \bar{\theta} &=& \bar{\psi} \bar{\theta} \theta \psi + c
\frac{i}{2} A^* \theta \overleftrightarrow{\partial} \cdot \sigma \bar{\theta}
A \label{Current}\\
&\equiv& \theta j \bar{\theta} + c \theta \Delta j \bar{\theta},
\end{eqnarray}
where
\begin{equation}
c=\frac{1}{1-k/2} \label{defineC}
\end{equation}
in accordance with the $R$ weights of the fields which are read off from
(\ref{Rsymmetry}). We have introduced grassman coordinates
$\theta^{\alpha},\bar{\theta}_{\dot{\alpha}}$
to contract and suppress spinorial indices. The dot implies a contraction
of Minkowski indices. 
The second term 
$(\Delta j)_{\alpha \dot{\alpha}}$ is known as the improvement term.

Let us check the supersymmetry transformation of (\ref{Current}), which 
acts as the lowest component of the supermultiplet;
\begin{eqnarray}
\theta \delta j \bar{\theta} &\equiv& \theta \eta^{\alpha} s_{\alpha}
\bar{\theta} + \theta \bar{\eta}_{\dot{\alpha}} \bar{s}^{\dot{\alpha}}
\bar{\theta} \nonumber \\
&=& \left( -i\sqrt{2} \eta \sigma \bar{\theta} \cdot \partial A^* +
\sqrt{2} \bar{\theta} \bar{\eta} {F}^* \right) \theta \psi
+\bar{\psi}\bar{\theta} \left( i\sqrt{2} \theta \sigma \bar{\eta} \cdot
\partial A+\sqrt{2}\theta \eta F \right), \label{supertrans1} \\
\theta \delta (\Delta j) \bar{\theta} &\equiv& \theta \eta^{\alpha}
(\Delta s)_{\alpha} \bar{\theta} + \theta \bar{\eta}_{\dot{\alpha}}
(\Delta \bar{s})^{\dot{\alpha}} \bar{\theta} \nonumber \\
&=& \frac{i}{2} A^* \theta \sigma \cdot \overleftrightarrow{\partial}
\bar{\theta} (\sqrt{2} \eta \psi) +
\frac{i}{2} (\sqrt{2} \bar{\eta} \bar{\psi}) \theta \sigma \cdot
\overleftrightarrow{\partial} \bar{\theta} A. \label{supertrans2}
\end{eqnarray}
The improved supercurrents are
\begin{equation}
-\frac{1}{2} \tr \bar{\sigma}^m \left( s_{\alpha}+c(\Delta s)_{\alpha} \right) \ \ \
\textrm{and} \ \ \ -\frac{1}{2} \tr \bar{\sigma}^m \left(
\bar{s}^{\dot{\alpha}}+c(\Delta \bar{s})^{\dot{\alpha}} \right) . \label{ISC}
\end{equation}
It is easy to check
\begin{equation}
\left. \theta \left( \delta j + c \delta (\Delta j) \right) \bar{\theta}
\ \right|_{\eta=\theta, \ \bar{\eta}=\bar{\theta}} =0
\end{equation}
if and only if $k=3$ and therefore $c=-2$ from (\ref{defineC}). This is
nothing but the condition that the supercurrents (\ref{ISC}) implement
the superconformal constraints, that is, the irreducibility of their spin 
when the coupling constant in the superpotential is dimensionless.

Let us further transform (\ref{supertrans1}) and (\ref{supertrans2}) to
generate the stress-energy tensor and we check the consistency with the
supersymmetry algebra as well;
\begin{eqnarray}
\theta \delta \delta j \bar{\theta} &=& (\bar{\psi}\bar{\theta})(\theta
\delta \delta \psi)+2(\delta \bar{\psi} \bar{\theta})(\theta \delta
\psi)+(\delta \delta \bar{\psi} \bar{\theta})(\theta \psi), \label{deldelj} \\
\theta \delta \delta (\Delta j) \bar{\theta} &=& \frac{i}{2} A^* \theta
\overleftrightarrow{\partial} \cdot \sigma \bar{\theta}(\delta \delta A)
+ i\delta A^* \theta \overleftrightarrow{\partial} \cdot \bar{\theta}
\delta A
+\frac{i}{2}(\delta \delta A^*) \theta \overleftrightarrow{\partial}
\cdot \sigma \bar{\theta} A\ . \label{deldelDelj}
\end{eqnarray}
The fermionic part of (\ref{deldelj}) is
\begin{eqnarray}
(\bar{\psi} \bar{\theta})(\theta \delta \delta \psi)+(\delta \delta
\bar{\psi} \bar{\theta})(\theta \psi)&=&-2i(\theta \eta)(\bar{\theta}
\bar{\eta})\bar{\psi}\bar{\sigma} \cdot \overleftrightarrow{\partial} \psi
+2i(\bar{\theta}\bar{\sigma}\theta) \cdot \left( (\bar{\psi} \bar{\eta})
\overleftrightarrow{\partial} (\psi \eta) \right) \nonumber \\
& &-2i(\bar{\eta}\bar{\sigma}\theta) \cdot \partial \left(
(\bar{\psi}\bar{\theta})(\eta \psi) \right)
+2i(\bar{\theta}\bar{\sigma}\eta) \cdot \partial \left(
(\bar{\psi}\bar{\eta})(\theta \psi) \right). \nonumber \\
\end{eqnarray}
The bosonic part of (\ref{deldelj}) is
\begin{eqnarray}
2(\delta \bar{\psi} \bar{\theta})(\theta \delta \psi) &=&
4(\eta\theta)(\bar{\eta} \bar{\theta})
 ({F}^*F - \partial A^* \cdot \partial A)
-4(\theta \sigma \bar{\theta}) \cdot \partial A^* 
 (\eta \sigma \bar{\eta}) \cdot \partial A
\label{bosonicof10}\\&&
+8(\eta \theta)(\bar{\theta} \bar{\sigma}^{mn} \bar{\eta})
 \partial_m A^* \partial_n A
-2i (\bar{\eta}\bar{\eta}){F}^*
 (\theta \sigma \bar{\theta}) \cdot \partial A
+2i(\eta \eta)(\theta \sigma \bar{\theta}) \cdot \partial A^* F \ .
\nonumber
\end{eqnarray}
The fermionic part of (\ref{deldelDelj}) is
\begin{equation}
i\delta A^* \theta \overleftrightarrow{\partial} \cdot \bar{\theta}
\delta A = 2i (\theta \sigma \bar{\theta}) \cdot (\bar{\eta} \bar{\psi}
\overleftrightarrow{\partial} \eta \psi).
\end{equation}
The bosonic part of (\ref{deldelDelj}) is
\begin{eqnarray}
&&\frac{i}{2} A^* \theta \overleftrightarrow{\partial} \cdot \sigma
\bar{\theta}(\delta \delta A)+\frac{i}{2}(\delta \delta A^*) \theta
\overleftrightarrow{\partial} \cdot \sigma \bar{\theta} A
\nonumber\\&&~~~~~
=-(\theta \sigma \bar{\theta} \cdot A^*
\overleftrightarrow{\partial})(\partial A \cdot \eta \sigma
\bar{\eta})+(\eta \sigma \bar{\eta} \cdot \partial
A^*)(\overleftrightarrow{\partial}A \cdot \theta \sigma \bar{\theta})
\nonumber \\&&~~~~~~~~~
+i(\eta \eta)\theta \sigma \bar{\theta} \cdot A^*
\overleftrightarrow{\partial}F+i(\bar{\eta} \bar{\eta})(\theta \sigma
\bar{\theta}) \cdot {F}^* \overleftrightarrow{\partial}A \ .
\label{bosonicof11}
\end{eqnarray}

The consistency of the supersymmetry algebra demands that the $\eta \eta$ term and the $\bar{\eta} \bar{\eta}$ term be absent in $\theta \delta \delta J \bar{\theta}$. 
Let us check that this is in fact the case. From (\ref{bosonicof10}) and (\ref{bosonicof11}), we see that the $\bar{\eta} \bar{\eta}$ term is 
\begin{eqnarray}
-2i (\bar{\eta} \bar{\eta}) F^* (\theta \sigma \bar{\theta}) \cdot \partial A + ci(\bar{\eta} \bar{\eta})(\theta \sigma \bar{\theta}) \cdot F^* \overleftrightarrow{\partial} A.
\end{eqnarray}
Using equation of motion for auxiliary fields $F,\ F^*$ and that $W(A)$ is a degree $k$ monomial in $A$, this is equal to
\begin{eqnarray}
2i(\bar{\eta} \bar{\eta}) \left( 1-\frac{c}{2} + \frac{c}{2}(k-1) \right) \theta \sigma \bar{\theta} \cdot \partial A, \label{app}
\end{eqnarray}
which vanishes when $c$ is chosen as (\ref{defineC}).

The remainder of $\theta \delta \delta J \bar{\theta}$ closes into the stress-energy tensor. Using equations of motion, we have checked 
\begin{eqnarray}
\theta \delta \delta J \bar{\theta}= -2 (c-1) T.
\end{eqnarray}
Here
\begin{eqnarray}
T &\equiv& \eta \sigma^m \bar{\eta} \theta \sigma^n \bar{\theta} T_{mn} \nonumber \\
  &=& -(\eta \sigma \bar{\eta}) \cdot \partial A^* (\theta \sigma \bar{\theta}) \cdot \partial A
  -(\eta \sigma \bar{\eta}) \cdot \partial A (\theta \sigma \bar{\theta}) \cdot \partial A^* \nonumber \\
  &&-\frac{i}{2} \left( \theta \sigma \bar{\theta} \cdot \bar{\eta} \bar{\psi} \overleftrightarrow{\partial} \eta \psi +\eta \sigma \bar{\eta} \cdot \bar{\theta} \bar{\psi} \overleftrightarrow{\partial} \theta \psi \right)
  +2 \eta \theta \bar{\eta} \bar{\theta} \mathcal{L}.
\end{eqnarray}


\bigskip


\end{document}